\documentclass[lettersize,journal]{IEEEtran}
\usepackage{amsmath,amsfonts}
\usepackage{algorithmic}
\usepackage{array}
\usepackage[caption=false,font=normalsize,labelfont=sf,textfont=sf]{subfig}
\usepackage{textcomp}
\usepackage{stfloats}
\usepackage{url}
\usepackage{verbatim}
\usepackage{graphicx}
\usepackage{color}
\usepackage{bm}
\usepackage{algorithm}

\newtheorem{definition}{Definition}

\newtheorem{lemma}{Lemma}
\newtheorem{corollary}{Corollary}
\newtheorem{theorem}{Theorem}

\def\BibTeX{{\rm B\kern-.05em{\sc i\kern-.025em b}\kern-.08em
    T\kern-.1667em\lower.7ex\hbox{E}\kern-.125emX}}
\usepackage{balance}
\begin{document}
\title{Gradient Flow Decoding 
\thanks{This work was supported by JSPS KAKENHI Grant Number JP22H00514.}
}

\markboth{Journal of \LaTeX\ Class Files,~Vol.~18, No.~9, September~2020}%
{How to Use the IEEEtran \LaTeX \ Templates}

\author{
\IEEEauthorblockN{Tadashi Wadayama}
\IEEEauthorblockA{ 
\textit{Nagoya Institute of Technology}\\
wadayama@nitech.ac.jp}
\and \\
\IEEEauthorblockN{Lantian Wei}
\IEEEauthorblockA{ 
\textit{Nagoya Institute of Technology}\\
wei.lantian@nitech.ac.jp}
}

\maketitle

\begin{abstract}
This paper presents the Gradient Flow (GF) decoding for LDPC codes. GF decoding, a continuous-time methodology based on gradient flow, employs a potential energy function associated with bipolar codewords of LDPC codes. The decoding process of the GF decoding is
	concisely defined by an ordinary differential equation
	and thus it is well suited to an analog circuit implementation.
 We experimentally demonstrate that the decoding performance of the GF decoding for AWGN channels
is comparable to that of the multi-bit mode gradient descent bit flipping algorithm.
We further introduce the negative log-likelihood function of the channel 
for generalizing the GF decoding.
The proposed method is shown to be tensor-computable, which means that 
the gradient of the objective function can be evaluated with the combination of 
basic tensor computations. 
This characteristic is well-suited to  emerging AI accelerators, potentially applicable in wireless signal processing. 
The paper assesses the decoding performance of the generalized GF decoding in LDPC-coded MIMO channels. Our numerical experiments reveal that the decoding performance rivals that of established techniques like MMSE + BP.
Furthermore, an exploration of score-based channel learning for capturing statistical properties is also provided.
\end{abstract}

\begin{IEEEkeywords}
LDPC codes, ordinary differential equations, gradient descent, 
tensor computation, MIMO channel, MIMO detection 
\end{IEEEkeywords}

\section{Introduction}

\subsection{Background}
Low-Density Parity-Check (LDPC) codes \cite{Gallager63} have emerged as a cornerstone 
in the ongoing advancement of contemporary wireless communication technologies. 
Their role is becoming increasingly critical, and
LDPC codes are expected to be necessary in ensuring reliable data transmission 
in the forthcoming 6G systems. 
The advent of 6G networks represents a substantial leap forward in communication technology, 
demanding levels of technical innovation.
For example, the seamless integration of Artificial Intelligence (AI) with signal processing 
stands as a pivotal challenge in the realization of 6G networks  \cite{Peltonen}. 
This integration is essential not only for enhancing network efficiency and capability 
but also for unlocking new potentials in wireless communications.


The recent remarkable achievements of Large Language Models (LLMs), 
such as GPT-4 \cite{OpenAI23}, 
have significantly influenced the technological landscape, particularly 
in the field of artificial intelligence. 
These successes have acted as a catalyst in the development of 
fast and power-efficient {\em AI accelerators} \cite{Reuther22}, 
including the next generation of Graphics Processing Units (GPUs). 
These advanced devices are specifically designed to cater to the demanding computational 
needs of AI models. They play a crucial role not only in accelerating 
the training process of these complex models but also in enhancing the efficiency 
of inference tasks. 

In recent years, optical integrated programmable circuits based 
on the Mach-Zehnder Interferometers (MZI) have garnered 
significant interest from researchers \cite{opt, uni}. 
An MZI is an optical component that consists of phase shifters, 
which alter the phase of the input light beams, 
and beam splitters, which divide it.
As an MZI has a two controllable parameters, it can be seen as a programmable optical component.
Optical matrix-vector product (MVP) circuit implemented using MZIs 
has been actively studied \cite{mzi1}. 
The programmable MZI-based MVP circuit is a  
promising candidate of the next generation AI accelerators because of 
its significant computation speed and power efficiency.
A recent example is a neural network implemented in the optical domain \cite{Zhang2021}. 


RRAM (resistive random-access memory)-based analog computing in electrical domain  
is another promising technology for analog computing.
Wang et al. \cite{Wang2023} recently presented 
a continuous-time resistive memory circuit for solving compressed 
sensing problem. In \cite{Zuo2023}, RRAM (resistive random-access memory)-based 
analog computing is applied for implementing MIMO precoding problems. It is demonstrated 
RRAM-based matrix-vector product circuits achieve high throughput and energy efficiency.


By offering significant enhancements in speed and energy efficiency, 
these next-generation AI accelerators could facilitate more extensive and practical applications 
across various fields, including wireless signal processing. 
Traditionally, a practical LDPC decoder is often implemented in an Application Specific 
Integrated Circuit (ASIC) for achieving sufficiently high throughput and energy efficiency.
However, the advancements in AI accelerators may be paving the way for the practical 
implementation of LDPC decoders on AI accelerators in the near future because 
{\em programability} provided by AI accelerators yields significant flexibility 
in wireless signal processing tasks handling various demands such as rate, latency, code type, and 
signal format.

For this direction, a next-generation decoding algorithm needs to be 'tensor-friendly',
i.e., each internal process of the decoding algorithm should be efficiently executable on 
an AI accelerator. This implies that each subprocess must be based on tensor computation such as 
	tensor product, tensor addition, 
	and component-wise function application, 
which is optimally managed by the AI accelerator. 
In other words, the core of algorithm should comprise tensor-friendly computations 
to fully leverage the power of the AI accelerator. 
A tensor-friendly decoding algorithm also facilitates batch based decoding, allowing for the simultaneous handling of multiple codewords. A batch-based algorithm promotes the efficient use of the AI 
	accelerator.
	
Furthermore, a tensor-friendly algorithm is a {\em differentiable algorithm}, i.e.,
every part of the algorithm is differentiable. This means that an automatic differentiation mechanism including back propagation can be easily and efficiently applied to compute the gradients of the internal trainable variables. We can embed a learnable block or learnable variables into 
the algorithm. For example, we can immediately use {\em deep unfolding} for performance 
improvement. Such a `AI-friendly' property is another benefit of a tensor-friendly algorithm.

\subsection{Contribution}

The objective of this study is {\em to devise a tensor-friendly decoding algorithm} for LDPC codes. 
To achieve the goal, we begin by a decoding approach 
based on a non-convex optimization problem.
Utilizing the 
gradient flow dynamics to minimize the objective 
function, we can derive a continuous time system 
for decoding LDPC codes.
This paper presents the Gradient Flow (GF) decoding for LDPC codes. GF decoding, a continuous-time methodology based on gradient flow, employs a potential energy function associated with bipolar codewords of LDPC codes.
This approach utilizes a potential energy function 
akin to the objective function employed in the Gradient Descent Bit Flipping (GDBF) algorithm \cite{wadayama10}. 

The key contributions of this paper are outlined as follows:
\begin{itemize}
	\item Introduction of the GF decoding method for AWGN channels.
	\item Demonstration that GF decoding is tensor-friendly, making it compatible with future AI accelerators.
	\item Expansion of GF decoding through the inclusion of the channel's negative log-likelihood function, enabling its application to a broad range of channels.
	\item  Development of a discretized version of GF decoding, analogous to a gradient descent method, tailored for digital AI accelerators such as  GPGPUs (General-Purpose computing on Graphics Processing Units).
	\item  Exploration on score-based channel learning for capturing statistical property.
\end{itemize}

The GF decoding can be regarded as the continuous-time counterpart of the GDBF algorithm since both approaches utilize fairly similar objective functions. The simplicity of the method enhances its suitability for implementation on AI accelerators. A notable feature of GF decoding is its {\em differentiability}, which allows for smooth integration with other AI components,
	such as a neural network mimicking a negative log likelihood of the target channel. 
	
\subsection{Outline}

The concept of GF decoding was initially presented in the conference papers \cite{wadayama23} 
and \cite{wadayama24}. The current paper enriches the initial presentation by offering a thorough exposition on the detailed derivation of GF decoding, an enhanced treatment of tensor-computability, and extensive experimental validations. Furthermore, this paper newly introduces the concept of score-based channel learning. These enhancements are anticipated to substantially bolster the understanding of the proposed method, offering deeper insights.

The structure of this paper is organized as follows:
Section II briefly reviews related works and introduces basic notation related to LDPC codes. Section III presents the idea of GF decoding for AWGN channels. Section IV discusses "tensor-computability," a rather simple computation model useful for AI accelerators. Section V broadens the scope of GF decoding to encompass general channels. Section VI shows the results of numerical experiments on MIMO channels. Section VII discusses score-based channel learning suitable for GF decoding. Finally, Section VIII provides a conclusive discussion.

\section{Preliminaries}

\subsection{Related works}

The method presented in this paper can be classified as part of the optimization-based decoding algorithms for LDPC codes.
 A number of works have been developed in this class.
The seminal work on Linear Programming (LP) decoding by Feldman \cite{Feldman03} clearly formulated the LDPC decoding problem as an LP problem.
Applications of interior point methods for solving the LP problem have been discussed in \cite{Vontobel08} \cite{Wadayama10b}.
A gradient descent formulation of a non-convex objective function 
leads to the GDBF algorithm \cite{wadayama10}, which introduced optimization perspective into 
the bit-flipping decoding algorithms.
Some variants of the GDBF algorithm, such as the noisy GDBF algorithm \cite{Sundararajan14}, 
have shown considerable improvement in decoding performance over a simple GDBF algorithm.
Zhang et al.\cite{Zhang13} and Barman et al.\cite{Barman13} applied 
Alternating Direction Method of Multipliers (ADMM) for solving the Feldman's LP problem,
which leads to ADMM decoding. ADMM decoding has been studied by many researchers and has become a decoding algorithm quite competitive with BP in both decoding performance and decoding complexity.

\subsection{Notation}

Assume that a sparse binary parity check matrix $\bm H = \{H_{ij}\} \in \mathbb{F}_2^{m \times n}$ 
is given.
The LDPC code $\tilde C(\bm H)$ is defined by 
\begin{align}
\tilde C(\bm H) \equiv \{\bm{b}  \in \mathbb{F}_2^n \mid  \bm H \bm{b} = \bm{0} \}.		
\end{align}
The binary to bipolar transform $\beta: \mathbb{F}_2 \rightarrow \{1, -1\}$
defined by $\beta(0) \equiv 1$ and $\beta(1) \equiv -1$ transforms 
$\tilde C(\bm H)$
into the bipolar code defined by
\begin{align}
C(\bm H) \equiv \{\beta(\bm{b}) \in \{1, -1\}^n \mid \bm{b}   \in \tilde C(\bm H) \}.		
\end{align}

The index sets $A(i)$ and $B(j)$ are defined as
\begin{align}
	A(i) &\equiv \{j \mid j \in [n], H_{i, j} = 1 \}, \quad i \in [m], \\
	B(j) &\equiv \{i \mid i \in [m], H_{i, j} = 1   \}, \quad j \in [n],
\end{align}
respectively.
The notation $[n]$ denotes the set of consecutive integers $\{1,2, \ldots, n\}$.
The parity check condition for the LDPC code $\tilde C(\bm H)$ is
\begin{align}
\forall \bm x \in \tilde C(\bm H), \quad \forall i \in [m],  \sum_{j \in A(i)} x_j  = 0.
\end{align}
This corresponds to the parity check condition for the bipolar LDPC code $C(\bm H)$, which is
\begin{align} \label{bipolar_pc}
\forall \bm x \in C(\bm H), \quad \forall i \in [m],  \prod_{j \in A(i)} x_j  = 1.
\end{align}

A function $g:\mathbb{C} \rightarrow \mathbb{C}$ can be applied to a vector $\bm x \in \mathbb{R}^n$ as
\begin{align}
g(\bm x) \equiv (g(x_1),g(x_2),\ldots, g(x_n)),
\end{align}
where $\bm x = (x_1,\ldots, x_n) \in \mathbb{C}^n$.
Namely, a scalar function $g$ can be component-wisely 
applicable to a vector $\bm x$.
For a pair of vectors $\bm a = (a_1,\ldots, a_n) \in \mathbb{C}^n, \bm b \equiv (b_1,\ldots, b_n) \in \mathbb{C}^n$, we define the multiplication and division on two vectors by 
$\bm a \bm b \equiv (a_1 b_1, \ldots, a_n b_n)$ and 
$\bm a / \bm b \equiv (a_1/b_1, \ldots, a_n/b_n)$, respectively.

\section{Gradient Flow Decoding for AWGN Channels}
\subsection{Gradient flow dynamics}

The gradient flow dynamics, also known as steepest descent dynamics, is a continuous-time dynamics defined by an Ordinary Differential Equation (ODE):
\begin{align} \label{flow}
	\frac{d\bm x(t)}{dt} = - \nabla f(\bm x(t)),
\end{align}
where $f:\mathbb{R}^n \rightarrow \mathbb{R}$ is a 
potential energy function. The system's state $\bm x:\mathbb{R} \rightarrow \mathbb{R}^n$
evolves to reduce the potential energy $f$ as the continuous time variable $t$ increases.
This continuous dynamical system can be viewed as the continuous counterpart of the gradient descent method.
 If the potential function $f$ is strictly convex, 
the equilibrium point of the dynamics coincides with the minimum point of $f$.
Therefore, the gradient flow dynamics can be seen as a continuous-time 
minimization process of the potential energy function.
If $f$ is a non-convex function, then
the gradient flow dynamics finds a stationary point of $f$ as $t \rightarrow \infty$.

\begin{figure}[htbp]
\begin{center}
\includegraphics[scale=0.35]{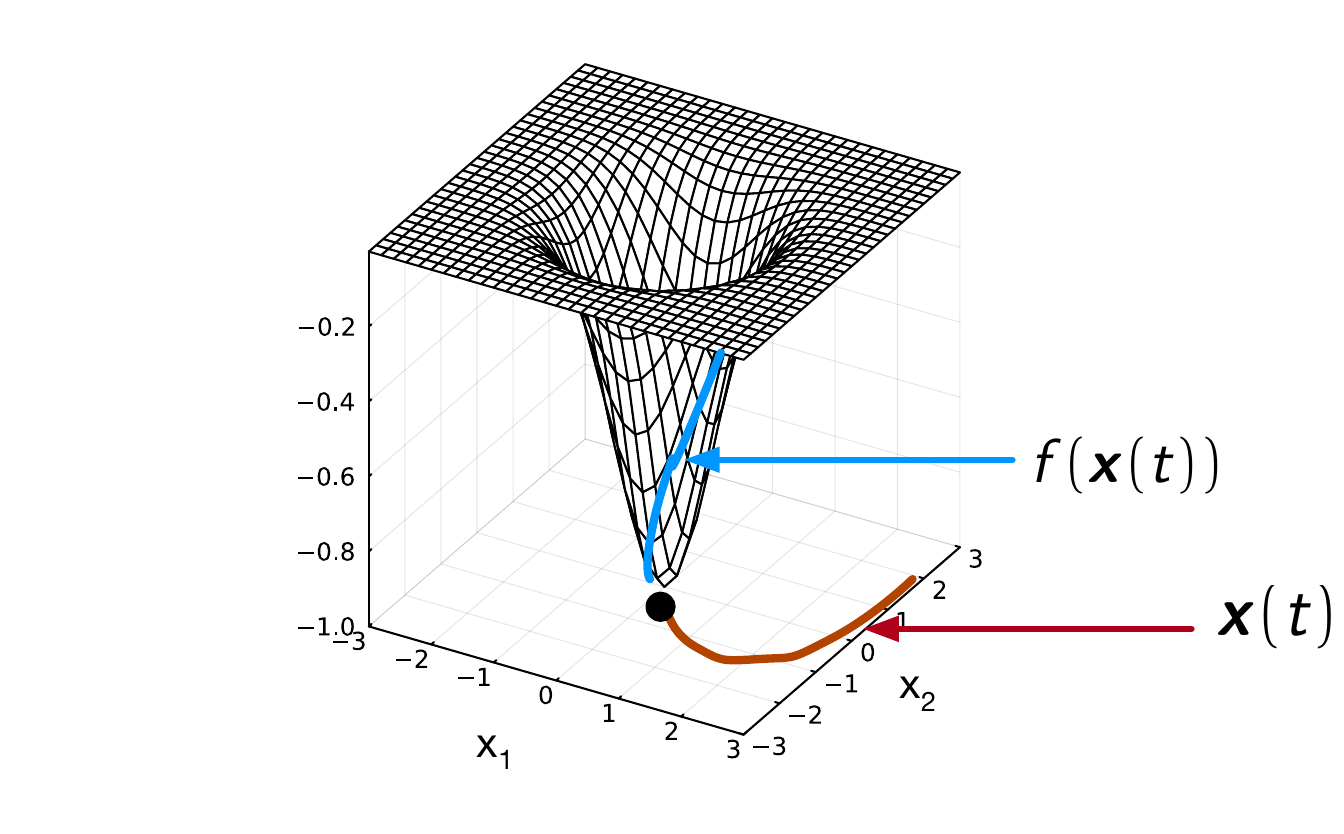}
\caption{Minimization process of a convex function in a gradient flow dynamics system: $\bm x(t)$ represents the solution to the ordinary differential equation (\ref{flow}). The equilibrium points become the minimum points of the convex function $f$.}
\label{GF}
\end{center}
\end{figure}

Figure \ref{GF} presents an example of the gradient flow dynamics for
a simple convex energy function ${\cal E}$ defined on two dimensional Euclidean space.
The state vector following the ODE (\ref{flow}) gradually approaches to the minimum point of ${\cal E}$.
The gradient flow can be considered as the continuous-time counterpart of the {gradient descent method}.

\subsection{Gradient flow decoding for AWGN channel}

We assume an AWGN channel.
A transmitter send a codeword $\bm s \in C(\bm H)$ and
a receiver obtain a received word 
\begin{align}
	\bm y = \bm s + \bm n,	
\end{align}
where $\bm n \sim {\cal N}(\bm 0, \sigma^2_n\bm I)$.

Let $f$ be a potential energy function defined by 
\begin{align} \label{potential_energy}
	f(\bm x) \equiv \frac 1 2 \|\bm x - \bm y\|^2 + h_{\alpha,\beta}(\bm x).
\end{align}
The first term of the potential energy function can be regarded as
the negative log likelihood function for an AWGN channel.
The second term is a penalty term attracting $\bm x$ to a codeword of $C(\bm H)$.

\begin{definition}
The {\em code potential energy function} for $C(\bm H)$ is a multivariate polynomial defined as
\begin{equation} \label{heq}
	h_{\alpha,\beta}(\bm{x}) \equiv \alpha \sum_{j = 1}^n (x_j^2 - 1)^2 
	+ \beta \sum_{i = 1}^m \left( \left(\prod_{j \in A(i)} x_j \right)  - 1 \right)^2, 
\end{equation}
where $\bm x = (x_1, \ldots, x_n)^T\in \mathbb{R}^n$.
The parameters $\alpha \in \mathbb{R}_+$ and $\beta \in \mathbb{R}_+$ control the 
relative strength of the first and second terms. \hfill\fbox{}
\end{definition}

The first term on the right-hand side of (\ref{heq}) 
represents the bipolar constraint for $\bm{x} \in \{+1, -1\}^n$,
and the second term corresponds to the parity constraint induced by $\bm H$, i.e.,
if $\bm x \in C(\bm H)$, we have
\begin{align}
\left(\prod_{j \in A(i)} x_j \right) -1 = 0	
\end{align}
for any $i \in [m]$.

The code potential energy $h_{\alpha,\beta}(\bm{x})$ is inspired by 
the non-convex parity constraint function used in the GDBF objective function \cite{wadayama10}.
The sum-of-squares form of (\ref{heq}) directly implies the most important property 
of $h_{\alpha,\beta}(\bm{x})$, i.e., 
the inequality 
\begin{align}
h_{\alpha,\beta}(\bm{x})  \ge 0	
\end{align}
holds for any $\bm{x} \in \mathbb{R}^n$. The equality holds if and only if $\bm{x} \in C(\bm H)$.

The gradient flow decoding for AWGN channel is simply a gradient flow dynamics based on 
the potential energy function $f$ that is given by
\begin{align}
	\frac{d\bm x(t)}{dt} = - \nabla f(\bm x(t)).	
\end{align}
This ODE can be rewritten as follows.
\begin{definition}[Gradient flow decoding for AWGN channels]
The GF decoding is defined by the ODE:
\begin{align}\label{GF_ODE1}
	\frac{d\bm x}{dt} &= - (\bm x - \bm y + \nabla h_{\alpha,\beta}(\bm x)) \\
	\bm x(0) &= \bm x_0,
\end{align}
where $h_{\alpha,\beta}(\bm x)$ is the code potential energy 
and  $\bm x(0) = \bm x_0 \in \mathbb{R}^n$ is an initial point. \hfill\fbox{}
\end{definition}

In general, there exist stationary points satisfying
\begin{align}
	f(\bm x) > 0,\quad \nabla f(\bm x) = \bm 0.
\end{align}
These stationary points are called {\em non-codeword stationary points}. Additionally, points that satisfy 
\begin{align}
f(\bm x) > 0, \nabla f(\bm x) \simeq \bm 0	
\end{align}
rather than equality are called {\em pseudo non-codeword stationary points}. When iterative optimization techniques that use gradient information, such as gradient methods, are applied, these non-codeword stationary points or pseudo non-codeword stationary points can cause the search vector's movement speed to drastically reduce near the stationary points, leading to traps at the stationary points.
These undesirable stationary points may negatively impact decoding performance because the search points may get trapped near them, preventing them from reaching the codeword stationary points.

\subsection{Analog circuit implementation}

\begin{figure}[htbp]
\begin{center}
\includegraphics[width=\columnwidth]{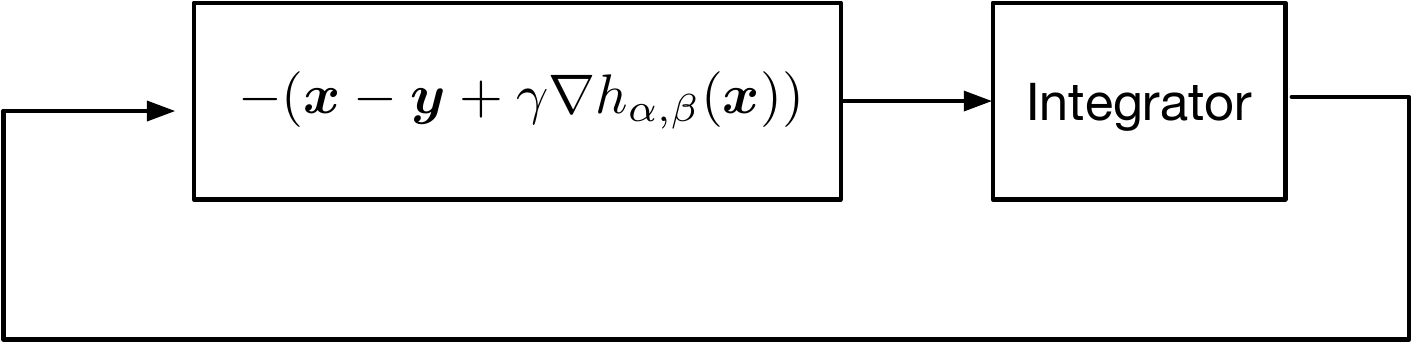}
\caption{Analog circuit for generalized GF decoding corresponding 
to the GF-ODE (\ref{GF_ODE1}).}
\label{GGF_decoding}
\end{center}
\end{figure}

The formulation of GF decoding is useful for continuous time signal processing by 
an analog circuit with feedback signals.
Figure \ref{GGF_decoding} presents a block diagram of the analog circuit realizing the 
continuous-time dynamics given by the generalized GF-ODE (\ref{GF_ODE1}).
The left box represents an analog circuit 
evaluating the value of $- (\bm x - \bm y + \gamma \nabla h_{\alpha,\beta}(\bm x))$.
A similar optical circuit implementing an Ising machine with feedback loop 
was reported in \cite{Prabhu20}. In principle, an optical implementation of the analog circuit 
depicted in Fig.\ref{GGF_decoding} would be possible if the gradient part 
$- (\bm x - \bm y + \gamma \nabla h_{\alpha,\beta}(\bm x))$ can be implemented 
with an optical circuit.

A tensor-friendly implementation of the circuit
evaluating $\nabla h_{\alpha,\beta}(\bm x)$ is to be discussed 
in Subsection \ref{grad_potential}.
The matrix-vector product in optical domain 
could be implemented with {\em programmable optical switches} 
when we use a quasi-cyclic LDPC codes.

\subsection{Euler method}
\label{Euler_sec}

To evaluate the decoding performance of the GF decoding (\ref{GF_ODE1}), a numerical method is required to solve the ODE. The simplest numerical method 
for solving simultaneous nonlinear differential equations is the Euler method~\cite{NumericalMethods}. 
While the convergence order of the Euler method is lower than that of higher-order methods, 
it is straightforward to use and can provide sufficiently accurate solutions 
if the time interval is sufficiently fine discretized. 
Therefore, in the present study, we will use the Euler method to solve the ODE (\ref{GF_ODE1}).

Suppose that we require to simulate the dynamics defined by the above ODE (\ref{GF_ODE1})
in the time interval $0 \le t \le T$.
The interval is divided into $N$ bins where $N$ denotes the number of discretized intervals.
The discrete-time ticks
$
t_{k} = k \eta (k = 0, 1, \ldots, N)
$
defines the boundaries of the bins 
where the width of a bin is given by 
$
\eta \equiv T/N.
$
It should be noted that the choice of the bin width  $\eta$ 
is crucial in order to ensure the stability and the accuracy of the Euler method. 
By using the Euler method,
the ODE (\ref{GF_ODE1}) can be approximated by 
\begin{align}
\bm x^{(k+1)} = \bm x^{(k)} - \eta \nabla f(\bm x^{(k)}) 	
\end{align}
for $k=0,1,2,\ldots, N-1$.
The initial condition of this recursion is $\bm x^{(0)} = \bm x_0$.

\subsection{Example of Code potential energy function}

As a simple example, suppose the potential energy function of the bipolar repetition code $C_{rep}$ ($n=2$) below. The code $C_{rep}$ is given by
\begin{align}
C_{rep} \equiv \{(1, 1), (-1, -1)\}.	
\end{align}
In this case, the corresponding potential energy function is
\begin{align}
h_{rep}(\bm{x}) = (x_1^2 - 1)^2 + (x_2^2 - 1)^2 + (x_1 x_2 - 1)^2,
\end{align}
where the constants $\alpha$ and $\beta$ are set to one.
It can be easily confirmed that by substituting the codewords into $h_{rep}(\bm{x})$, the potential energy value becomes $0$, and for any $\bm x$, $h_{rep}(\bm{x}) \ge 0$.

\begin{figure}[t]
\begin{center}
\includegraphics[scale=0.35]{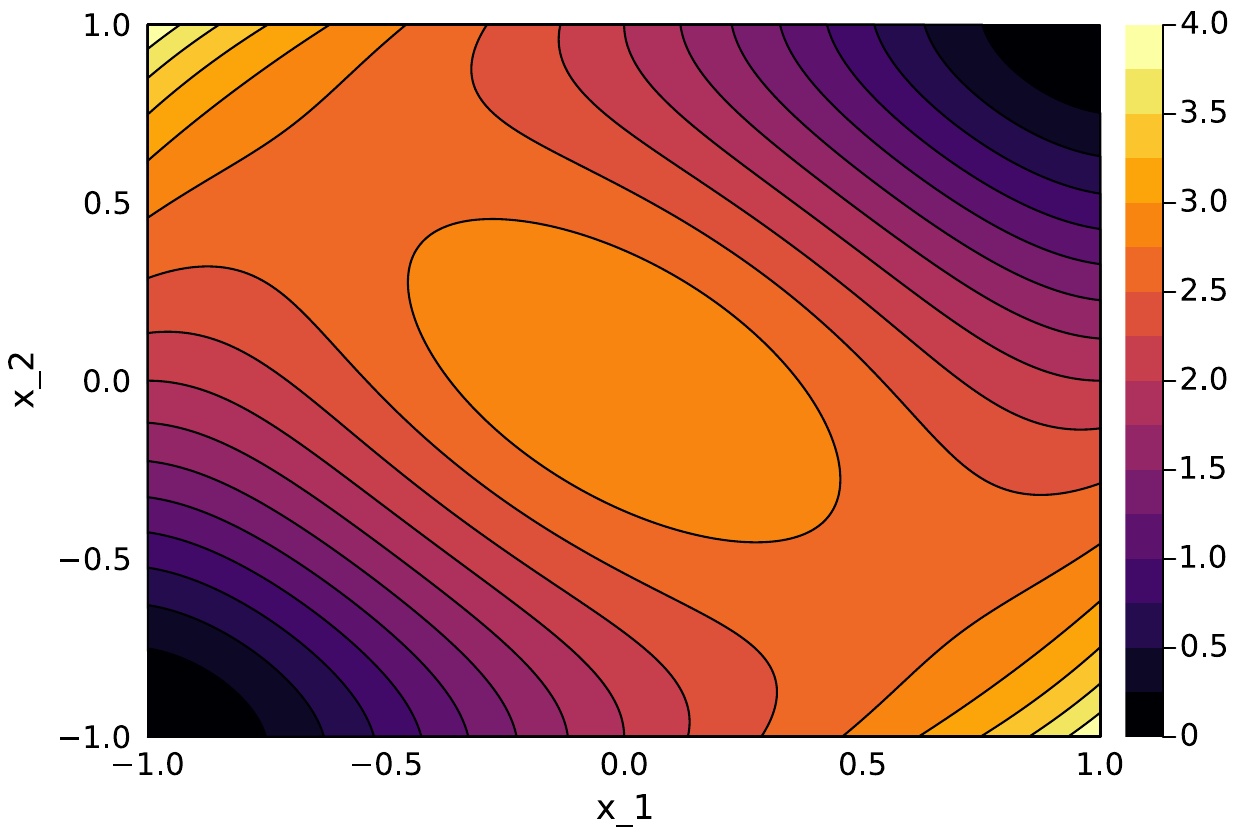}		
\end{center}
\caption{Heat map of $h_{rep}(\bm{x})$.}
\label{hrep_contour}
\end{figure}

The heatmap representation of the potential energy function $h_{rep}(\bm{x})$ is shown in Fig.~\ref{hrep_contour}. From this figure, it can be seen that the function takes a minimum value of $0$ at the locations of the codewords $(+1, +1), (-1, -1)$. The shape of the function $h_{rep}(\bm{x})$ resembles a valley with these two codewords at its base. Therefore, it is expected that when optimization techniques such as gradient methods are used, 
a state vector is attracted towards the direction of the codewords will occur. On the other hand, the origin $(0,0)$ is a maximum point, and $\nabla h_{rep}(\bm 0) = \bm 0$ holds. Therefore, if a state vector following the gradient descent dynamics passes near the origin, it is expected that its movement speed will be extremely slow.

\subsection{Example of GF Decoding Process}

We first present a small example 
to illustrate a decoding process. 
Suppose that we have a repetition code of length 2, $C_{rep}$.
Assume that a transmitted word is $\bm x = (1,1)$ and 
that the corresponding received word is $\bm y = (0.6027, 0.8244)$.
In this case, the ODE (\ref{GF_ODE1}) for the GF decoding becomes
\begin{align} \nonumber
&\left(
	\begin{matrix}
	\frac{dx_1}{dt} \\
	\frac{dx_2}{dt}
	\end{matrix}
\right)	 
= 
-\left(
	\begin{matrix}
	x_1 - 0.6027 + 4x_1(x_1^2 - 1) + 2 (x_1 x_2-1)x_2 \\
	x_2 - 0.8244 + 4x_2(x_2^2 - 1) + 2 (x_1 x_2-1)x_1
	\end{matrix}
\right).	
\end{align}

\begin{figure}[htbp]
\begin{center}
\includegraphics[scale=0.35]{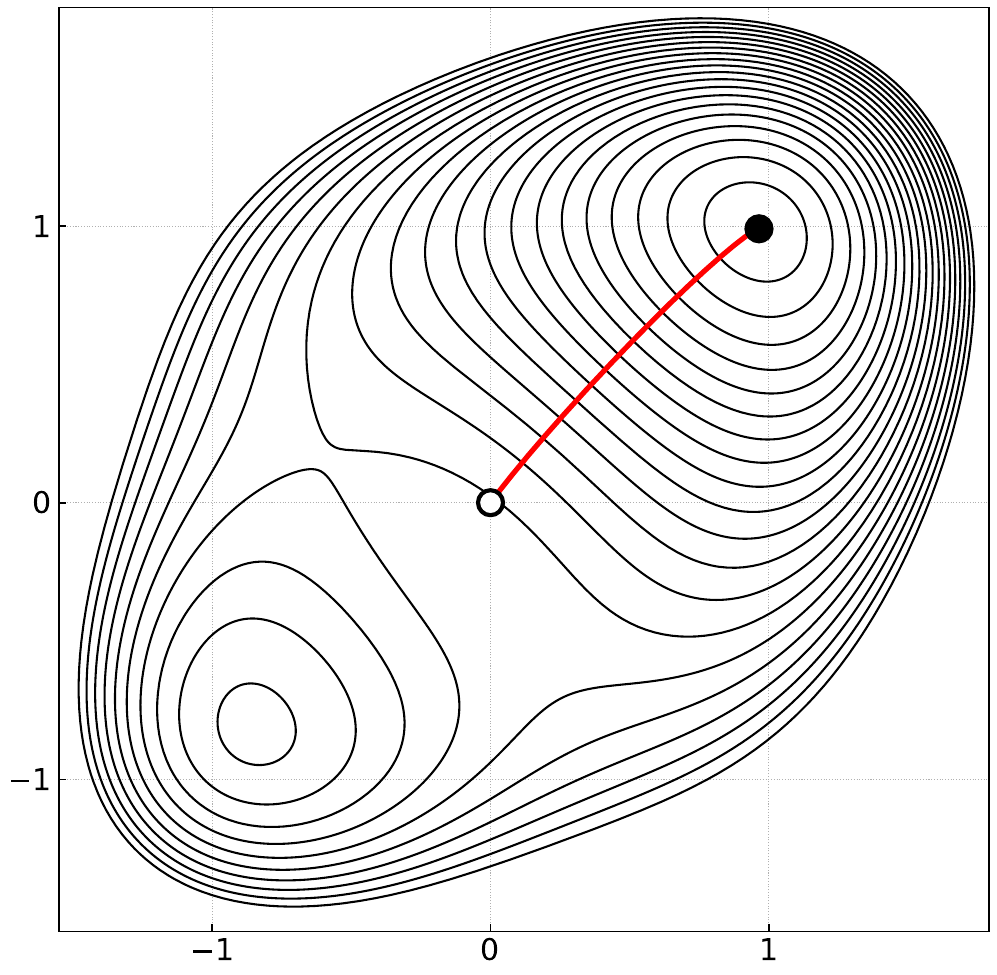}
\caption{Example of solution curve. The repetition code of length 2 is assumed.}
\label{fig:repetition}
\end{center}
\end{figure}

The initial condition for the ODE is set to $\bm x(0) = (0,0)$. Figure \ref{fig:repetition} shows the solution curve of the ODE, with the white small circle representing the initial point and the black small circle indicating the equilibrium point $(0.9642, 0.9901)$. The solution curve is obtained numerically using the Euler method. In Fig.~\ref{fig:repetition}, the contour curves of the potential energy function $f(\bm x)$ are also plotted. We can observe that the solution curve is orthogonal to the contour curves because the solution path follows the negative gradient vector field of the potential energy. This means that the curve is a steepest descent curve for $f$. By rounding the equilibrium point, we obtain $\hat{\bm x} = (1,1)$, which is the correct estimated word.

\begin{figure}[htbp]
\begin{center}
\includegraphics[scale=0.4]{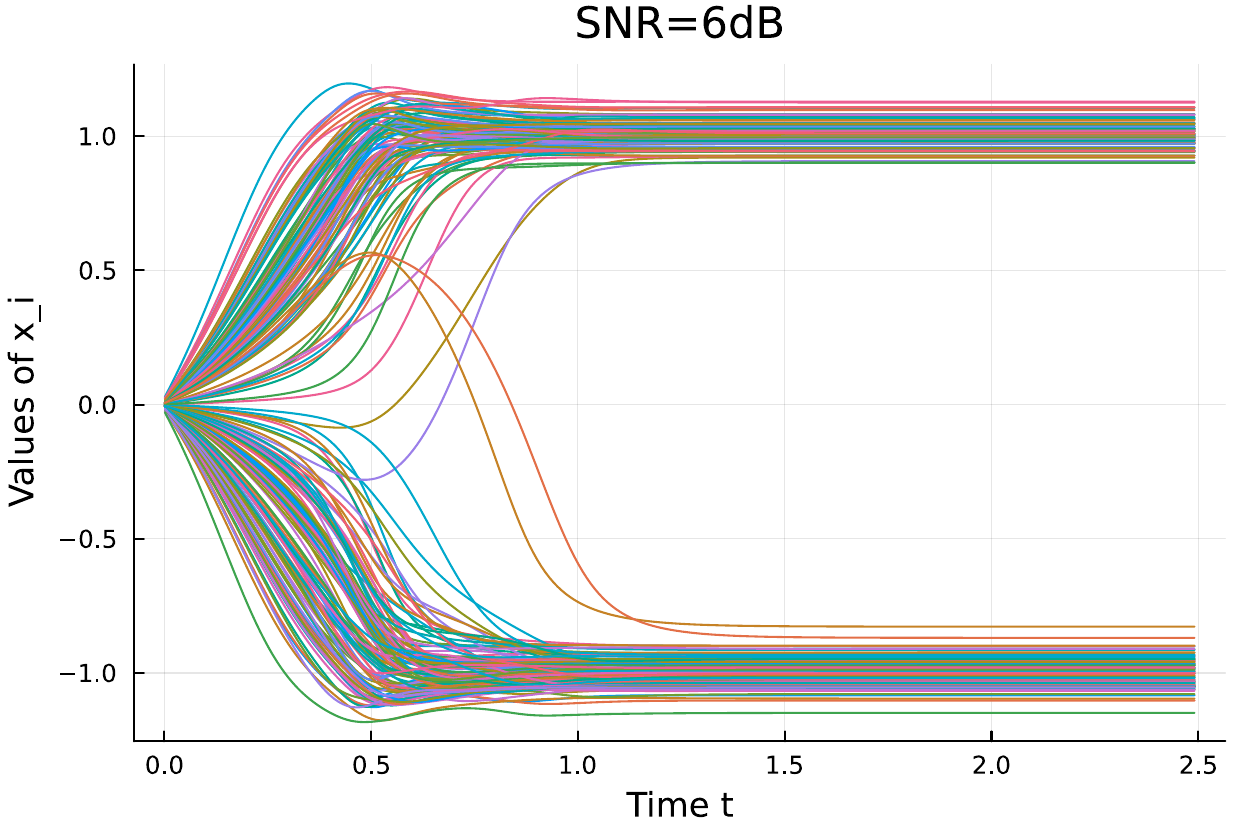}
\caption{Solution curves. (3,6)-regular LDPC codes of length 204.}
\label{fig:trajectory}
\end{center}
\end{figure}

We then show solution curves for an LDPC code.
Figure~\ref{fig:trajectory} displays the solution curves of each element in the state vector $\bm x$, where $x_i(t)$ (for $i \in [204]$) represents the solution of the ODE (\ref{GF_ODE1}) plotted as a function of time $t$. Notably, all solution curves are smooth and continuous. 
It can be seen that the equilibrium point is in proximity to a bipolar vector. 

\subsection{BER Performance of GF decoding}

We evaluated the bit error rate (BER) of GF decoding through computer simulations on LDPC codes with design rate 1/2, the $(3,6)$-regular LDPC codes (96.33.964, 204.33.484, PEGReg252x504, PEGReg504x1008) \cite{MacKay}. We used BP decoding as the baseline and set the maximum iteration of BP to 100. The parameter setting 
of GF decoding is summarized in Table \ref{parameters}.

\begin{figure}[htbp]
\begin{center}
\includegraphics[width=\columnwidth]{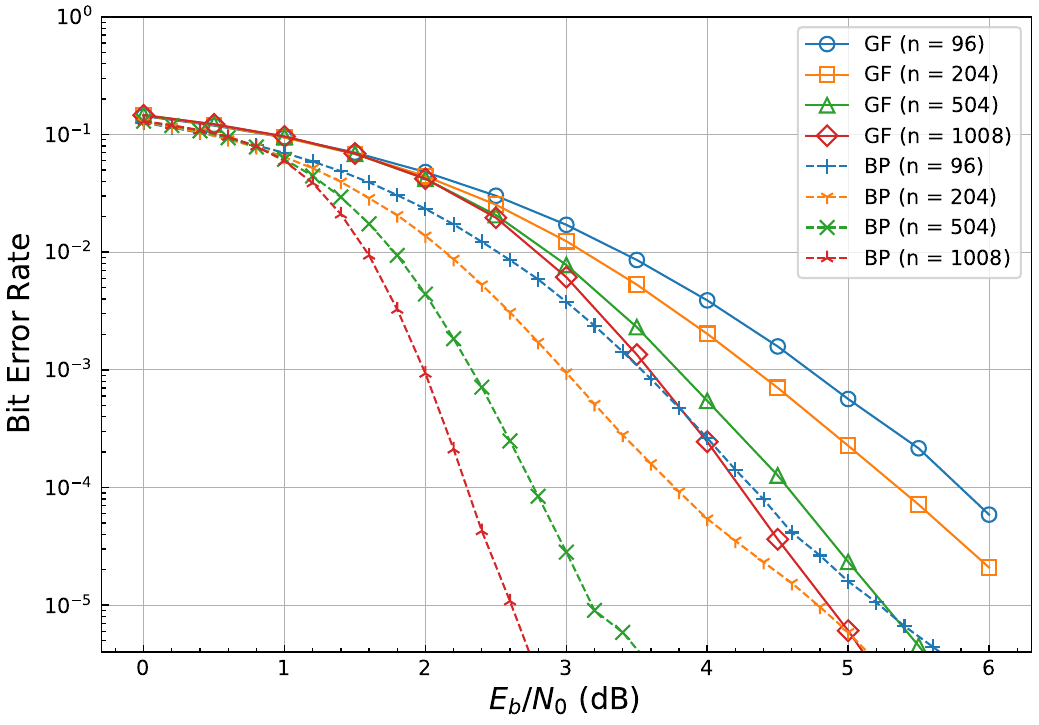}
\caption{Bit error rate (BER) of the GF decoding for AWGN channels. The BERs of BP are also included as benchmarks.}
\label{fig:BER}
\end{center}
\end{figure}

Figure \ref{fig:BER} displays the BER performance of the proposed GF decoding. Compared to BP, the GF decoding has a BER performance that is approximately 2dB worse. Notably, for PEGReg504x1008, the GF decoding performance is almost on par with the multi-mode GDBF algorithm using 100 iterations (\cite{wadayama10}, Fig.~3). Overall, GF decoding's BER performance is comparable to that of bit flip-type decoding algorithms.

\begin{table}[htbp]
\caption{Parameter setting of GF decoding.}
\label{parameters}
\centering{
\begin{tabular}{ll}
	\hline
	\hline
Potential energy parameters &	$\alpha=1, \beta=2$ \\
Parameters of Euler method   & $T = 10, N = 1000$ \\
Sampling time & $t = 10$ \\
Encoding & Uniformly random codeword \\
\hline
\end{tabular}}
\end{table}

\section{Tensor-computability}

\subsection{Computation model}
Our target hardware includes AI accelerators, such as GPGPUs, 
and programmable optical circuits.
Such hardware is especially efficient for tensor-based computations.
Here, we introduce the concept of tensor-computability to clarify 
our computation model.
\begin{definition}
The function $f:\mathbb{C}^n \rightarrow \mathbb{C}^n$ is said to be {\em tensor-computable} if 
the value of $f$ can be evaluated by a combination of the following basic operations:
\begin{itemize}
	\item Tensor product, e.g., matrix-vector product.
	\item Scalar multiplication, e.g., $\alpha \bm x (\alpha \in \mathbb{C},\ \bm x \in \mathbb{C}^n)$.
	\item Component-wise function applications to a tensor 
	\item Vector addition and subtraction, e.g., $\bm a + \bm b, \bm a - \bm b$.
\end{itemize}
\end{definition}


A 'tensor-computable' function can be naturally evaluated  with parallel computation because its basic operations  inherently has parallelism.
This concept of tensor-computability could also be viewed as the essential criteria for basic, efficient operations that a future AI accelerator should support. While contemporary AI accelerators have the capacity to handle more complex tensor operations, we restrict the permissible operations to maintain a straightforward and clear definition of tensor-computability.

An additional advantage of tensor-computable functions is their compatibility with {\em batch processing}, crucial for fully leveraging the computational power of GPGPUs. Moreover, this batch processing aligns seamlessly with contemporary neural network frameworks like PyTorch, TensorFlow, and JAX, 
indicating that tensor-computable functions are well-suited for AI tasks. In essence, a tensor-computable function is highly AI-friendly as well.

In the realm of optical computing, 
MZI-based Matrix-Vector Product (MVP) 
circuits have already been implemented \cite{uni}, 
and the computation of logarithmic and exponential functions in the optical domain 
has been explored \cite{Jiang16}. 
Optical switches and crossbar can be used for 
constructing a programmable MVP circuit
with a permutation matrix.
An implementation of RRAM-based analog matrix computing is discussed in \cite{Zuo2023}.
We may be able to implement a tensor-computable function with such optical or electrical devises,
which provides high processing speed and energy efficiency.

\subsection{Tensor computability of Gradient of code potential energy}
\label{grad_potential}

For implementing the GF decoding, we need to evaluate the gradient of the code potential energy.
The first order derivative of the potential energy can be easily evaluated as 
\begin{align} \nonumber
	&\frac{\partial}{\partial x_k}h_{\alpha,\beta}(\bm{x}) \\ \nonumber
	&= 4\alpha(x_k - 1)(x_k + 1) x_k \\ \label{gradient_expression}
	&+ 2\beta\sum_{i \in B(k) } \left( \left(\prod_{j \in A(i)} x_j \right)  - 1 \right) 
	\left(\prod_{j \in A(i) \backslash \{ k \} } x_j \right). 
\end{align}	
The gradient $\nabla h_{\alpha,\beta}(\bm{x})$ is thus given by
\begin{equation}
	\nabla h_{\alpha,\beta}(\bm{x}) \equiv \left(\frac{\partial}{\partial x_1}h_{\alpha,\beta}(\bm{x}), \ldots, \frac{\partial}{\partial x_n}h_{\alpha,\beta}(\bm{x})     \right)^T.
\end{equation}

The following lemma indicates tensor-computability of the gradient of the code potential energy.
\begin{lemma} \label{tensor_computability}
The gradient $\nabla h_{\alpha,\beta}(\bm{x})$ can be evaluated by
\begin{align} 
\nabla h_{\alpha,\beta}(\bm{x}) &= 4 \alpha \exp\left( \bm z_{-1} +  \bm z  +  \bm z_{+1} \right)
+ 2 \beta \exp \left(\bm w - \bm z  \right),
\end{align}
where $\bm z, \bm z_{-1}, \bm z_{+1}, \bm w$ are defined by
\begin{align} \label{log_func}
\bm z &\equiv \ln(\bm x), \\
\bm z_{-1} &\equiv \ln(\bm x - \bm 1), \\
\bm z_{+1} &\equiv \ln(\bm x + \bm 1), \\
\bm w &\equiv \ln (\bm H^T \left(\exp(2 \bm H \bm z)-\exp(\bm H \bm z) \right) )
\end{align}
if $\bm x \ne \bm 0$.
This implies that $\nabla h_{\alpha,\beta}(\bm{x})$ is tensor-computable.
\end{lemma}
(Proof)
From Eq.(\ref{log_func}), we have 
$
x_i	= \exp(z_i),
$
and it results in
\begin{align}
\prod_{j \in A(i)} x_j 
&= \prod_{j \in A(i)} \exp(z_i) \\
&= \exp\left(\sum_{j \in A(i)}z_i \right).
\end{align}
It is evident that the following equality holds:
\begin{align} \label{Q_eq}
\left(
\begin{array}{c}
Q_1 \\	
Q_2 \\	
\vdots \\
Q_m \\	
\end{array}
\right) \equiv \exp(\bm H \bm z),
\end{align}
where $Q_i (i \in [m])$ is defined by
\begin{align}
	Q_i \equiv \left(\prod_{j \in A(i)} x_j \right).	
\end{align}
Note that the binary matrix $\bm H$ is originally defined over $\mathbb{F}_2$ 
but the matrix $\bm H$ is now embedded into $\mathbb{R}$. 
Namely, the matrix vector product $\bm H \bm z$ is carried out over $\mathbb{R}$.
By using (\ref{Q_eq}), we have 
\begin{align}
	& 2\beta\sum_{i \in B(k) } \left( \left(\prod_{j \in A(i)} x_j \right)  - 1 \right) 
	\left(\prod_{j \in A(i) \backslash \{ k \} } x_j \right) \\
	&= 	\frac{2 \beta}{x_k} \sum_{i \in B(k) }\left( Q_i^2  - Q_i \right) \\ \label{w_expression}
	&= \frac{2 \beta}{x_k} \left[\bm H^T (\exp(2\bm H \bm z) - \exp(\bm H \bm z)) \right]_k,
\end{align}
where the notation $[\cdot]_k$ represents the $k$th component of the vector.
We thus have 
\begin{align}
	 4 \alpha(\bm x - \bm 1)(\bm x + \bm 1)\bm x
	 = 4 \alpha \exp\left(\bm z_{-1} +  \bm z  +  \bm z_{+1} \right),
\end{align}
\begin{align}
2\beta\frac{\exp(\bm w)}{\bm x} = 2 \beta \exp \left(\bm w - \bm z  \right),
\end{align}
which yields the claim of the lemma. \hfill \fbox{}

It should be noted that 
the evaluation method presented in Lemma \ref{tensor_computability} may not be 
optimal in terms of the computing time.
The computing time for evaluating 
$\nabla h_{\alpha,\beta}(\bm{x})$ could be improved 
depending on the basic tensor operations
provided by an AI accelerator.
For example, some AI accelerators have
efficient sparse MVP operation which can 
make use of the sparseness of the parity check matrix $\bm H$.
While the evaluation approach detailed in Lemma \ref{tensor_computability} may not be the most efficient for some AI accelerators, Lemma \ref{tensor_computability} clearly demonstrates that the process of evaluating gradients can be effectively parallelized.

\subsection{Thread computation model for GPGPU}
\label{thread_model}

In estimating the computational complexity 
and required computational resource with the assumption of GPGPU, 
a model that takes into account {\em thread parallelism} is necessary. 
Below is a simple computational model that considers thread parallelism for evaluating the gradient of code potential 
	energy function $\nabla h_{\alpha,\beta}(\bm x)$.

Our basic assumptions are as follows:
\begin{itemize}
\item Assuming {batch processing} with a batch size of $D$.
\item The size of $\bm H$ is $m \times n$.
\item A single thread can handle up to $O(n)$ processing.
\end{itemize}

%

Evaluating the gradient of the code potential energy necessitates computing the matrix-matrix multiplication $\bm H \bm Z$, where $\bm Z \in \mathbb{C}^{n \times D}$ represents a batch of state vectors. This step is the most demanding part of the gradient evaluation process as described in Lemma \ref{tensor_computability}. It is assumed that each individual thread will handle the computation of the dot product between a row vector from $\bm H$ and a column vector from $\bm Z$. This dot product operation is carried out sequentially by each thread, and the computational load per thread is $O(n)$. In such a scenario, $m D$ threads are needed to work concurrently in computing $\bm H \bm Z$. Thus, the parallel computation of the gradient evaluation {\em necessitates the use of $m D$ threads}.
	This is the dominant use of the computational resources 
	of a GPGPU.

A comparison of the required computational resources between GF decoding and tensor-computable BP decoding is presented in the Appendix. GF decoding can be seen as a lighter algorithm that requires fewer GPU computational resources because it involves matrix computations with smaller size matrices.

\section{Generalization of Gradient flow decoding}

\subsection{Approximate maximum a posteriori (MAP) decoding}

The GF decoding is defined only for AWGN channel in the previous sections.
In this section, we will generalize GF decoding for a general vector channel
defined by a conditional probability distribution $p(\bm y | \bm x)$.

We here briefly review the approximate MAP decoding introduced in the paper on 
{proximal decoding} \cite{wadayama23b}.
Assume that a sender transmits a codeword of $C(\bm H)$ to a given channel.
The channel is defined by a probability density function (PDF),
$p(\bm y |\bm x) (\bm x\in \mathbb{R}^n, \bm y \in \mathbb{R}^{N})$.
The negative log-likelihood function is defined as
\begin{align}
L(\bm x;\bm y) \equiv -\ln p(\bm{y}|\bm{x}).			
\end{align}
The optimal decoding rule for this channel is MAP decoding rule  defined as
\begin{align}
\hat{\bm x} \equiv\text{argmax}_{\bm x \in \mathbb{R}^n}\ 	p(\bm y| \bm x) p(\bm{x}),	
\end{align}
where $p(\bm{x})$ is the prior distribution. In general, 
MAP decoding is computationally intractable and we thus need to consider an 
approximation of MAP decoding for sufficiently long codes.

The ideal prior distribution is based on the code $C(\bm H)$ which is given by
\begin{align}
p(\bm{x}) \equiv \frac{1}{|C(\bm H)|}\sum_{\bm c \in C(\bm H)}\delta(\bm x - \bm c),		
\end{align}
where $\delta$ is Dirac's delta function.
We here introduce a Gibbs prior distribution
\begin{align}
\tilde p(\bm{x}) \equiv \frac{1}{Z} \exp\left(-\gamma h_{\alpha,\beta}(\bm x) \right),			
\end{align}
where $Z$ is the normalizing constant and $\gamma$ is a positive constant.
The function $h_{\alpha,\beta}(\bm x)$ takes the value zero if and only if $\bm x \in C(\bm H)$
and takes a positive value otherwise.
This implies that we have
\begin{align}
\tilde p(\bm{x}) = \frac 1 Z \exp\left(-\gamma h_{\alpha,\beta}(\bm x) \right) \rightarrow \frac{1}{|C(\bm H)|}\sum_{\bm c \in C(\bm H)}\delta(\bm x - \bm c)
\end{align}
at the limit $\gamma \rightarrow \infty$. 
This fact suggests the following approximation on the posterior PDF:
\begin{align} \nonumber
p(\bm x| \bm y) &\propto p(\bm y| \bm x) p(\bm x) 
\simeq 	p(\bm y| \bm x) \tilde p(\bm x) \\
&= \exp\left(- L(\bm x;\bm y) - \gamma h_{\alpha,\beta}(\bm x) \right).
\end{align}
From this approximate posterior PDF, we can 
define a following decoding rule.
\begin{definition}
The approximate MAP rule considered here is given by
\begin{align}
\hat{\bm x} \equiv \text{argmin}_{\bm x \in \mathbb{R}^n}\ 	\left[L(\bm x;\bm y )  +  \gamma h_{\alpha,\beta}(\bm x) \right],
\end{align}
where $\gamma$ is a positive real number.
\hfill\fbox{}	
\end{definition}

\subsection{Generalization of GF decoding}

In the previous subsection, the concept of an approximate MAP rule was introduced.
The objective function to be minimized is 
\begin{align}
F(\bm x) \equiv L(\bm x;\bm y ) + \gamma  h_{\alpha,\beta}(\bm x).	
\end{align}
There are many possibilities for solving the optimization problem.
For example, proximal decoding aims to solve this continuous minimization problem by using a proximal gradient descent method \cite{wadayama23b}.
In the subsequent argument, we will solve this minimization problem by using the continuous-time gradient flow dynamics.
The definition of the generalized GF decoding \cite{wadayama24} is given as follows.
\begin{definition}[Generalized GF decoding]
The ODE for the generalized GF decoding is given by 
\begin{align}\label{General_ODE}
	\frac{d\bm x}{dt} &= - (\nabla L(\bm x;\bm y ) + \gamma \nabla h_{\alpha,\beta}(\bm x)), 
\end{align}
where the initial condition is $\bm x(0) = \bm x_0$.
\hfill\fbox{}
\end{definition}
For example, for AWGN channels, the objective function has the form: 
\begin{align}
	F_{\rm{AWGN}}(\bm x) \equiv \frac 1 2 \|\bm x - \bm y\|^2 +  \gamma h_{\alpha,\beta}(\bm x),
\end{align}
where $\bm y$ represents the received word.
This means that the generalized GF-ODE (\ref{General_ODE})
becomes identical to the original GF-ODE (\ref{GF_ODE1}).

If the righthand side of the generalized GF-ODE (\ref{General_ODE}),
\begin{align}
\nabla F(\bm x) = \nabla L(\bm x;\bm y ) + \gamma \nabla h_{\alpha,\beta}(\bm x),
\end{align}
is tensor-computable, the whole decoding processes can be efficiently executed in an
AI accelerator. 
The following theorem clarifies the tensor-computability of $\nabla F(\bm x)$.

\begin{theorem}
If $\nabla L(\bm x;\bm y )$ is tensor-computable, then 
$\nabla F(\bm x)$ is tensor-computable.
\end{theorem}
(Proof) To prove the claim of theorem, 
it is sufficient to show that 
$\nabla h_{\alpha,\beta}(\bm x)$ is tensor-computable. Lemma \ref{tensor_computability}  
establishes the tensor-computability of the gradient of code potential energy funciton. \hfill \fbox{}

\subsection{Discrete-time version of GF decoding}


When implementing GF decoding on a GPGPU, it is necessary to discretize the continuous GF decoding process. The Euler method can be employed to discretize the minimization processes of the generalized GF-ODE (\ref{General_ODE}). This approach yields the recurrence formula for a discrete-time version of GF decoding:
\begin{definition}[Discretized GF decoding]
The recursive formula of the discretized GF decoding
is given by 
\begin{align} \label{GFGD}
	\bm s^{(k+1)} &= \bm s^{(k)} - \eta (\nabla L(\bm s^{(k)};\bm y ) + \gamma \nabla h_{\alpha,\beta}(\bm s^{(k)})),\ k = 0,1,\ldots,
\end{align}
where $\bm s^{(0)} = \bm x_0$. \hfill\fbox{}
\end{definition}
The iterative process in (\ref{GFGD}) can be regarded as a 
gradient descent method for minimizing $F(\bm x)$. 

We also refer 
to the discretized decoding process (\ref{GFGD}) as GF decoding
in this paper.
GPGPUs and digital AI accelerators 
can be used to execute the discrete time 
iterative process defined in (\ref{GFGD}).
In such cases, computation of 
$\nabla L(\bm s^{(k)};\bm y )$ 
should be tensor-computable, 
as these devices are designed for tensor-friendly computations.

\subsection{Avoiding Numerical Instability}

The discretized GF decoding described above can become numerically unstable when the step size parameter $\eta$ is relatively large. 

Let $\xi$ be a positive constant slightly larger than 1, and let 
$
B_{\xi} \equiv [-\xi, \xi]^n 	
$
represent an $n$-dimensional hypercube. Here, 
$
[a,b] \equiv \{x \in \mathbb{R} \mid a \le x \le b\}.
$
The norm of the gradient $\|\nabla h_{\alpha,\beta}(\bm{x})\|$ tends to become very large when $\bm x \notin B_{\xi}$ due to the properties of the potential energy function, which can cause numerical instability (oscillations or divergent behavior) in the aforementioned neighborhood decoding process. In such cases, to prevent numerical instability, instead of (\ref{GFGD}), one can use
\begin{equation}\label{box_proj}
\bm{s}^{(k+1)} = \Pi_\xi\left(\bm s^{(k)} - \eta (\nabla L(\bm s^{(k)};\bm y ) + \gamma \nabla h_{\alpha,\beta}(\bm s^{(k)}))  \right).
\end{equation}
The projection operator $\Pi_\xi: \mathbb{R}^n \rightarrow B_{\xi}$ represents the projection onto $B_{\eta}$, and is defined as
\begin{align}
	\Pi_\xi(\bm x) \equiv \arg \min_{\bm x' \in B_{\xi}} \|\bm x - \bm x'\|.	
\end{align}
It is experimentally confirmed that this projection operation can stabilize the decoding process.

\subsection{Deep Unfolding}


The recursive equation (\ref{GFGD}) incorporates multiple internal parameters, such as $\eta$, $\gamma$, $\alpha$, and $\beta$. It has been confirmed through experiments that the choice of these parameters significantly  influences the decoding performance. To fine-tune these parameters, several strategies can be employed, such as conducting preliminary experimental searches, employing random or grid searches, and using Bayesian optimization for hyper-parameter optimization. Nonetheless, pinpointing the ideal parameter setup is a computationally intractable task due to the extensive number of tunable parameters.

{\em Deep unfolding} is a method that seeks to boost the performance 
of iterative algorithms by optimizing the internal parameters via 
deep learning techniques.
In this approach, 
iterative processes are unfolded into the time direction
and each iteration is treated as a layer of 
a deep neural network.
For example, many sparse signal recovery algorithms and 
MIMO detection algorithms 
are iterative in nature, 
applying deep unfolding can lead to significant improvements 
in both recovery performance and convergence speed \cite{LISTA, 9020494,8759948,TISTA, He2020}.

In the following discussion, we will study 
the GF decoding algorithm with the set of 
internal trainable parameters:
\begin{align} \nonumber
 \{\eta^{(k)}, \gamma^{(k)}, \alpha^{(k)}, \beta^{(k)} \} 	
\end{align}
 for $k =1,2,\ldots, T$. 
Namely, the recursive formula for the GF decoding becomes as follows.
\begin{definition}[GF decoding with trainable parameters]
\begin{align} \label{GFGD_trainable}
	\bm s^{(k+1)} &= \bm s^{(k)} - \eta^{(k)} (\nabla L(\bm s^{(k)};\bm y ) + \gamma^{(k)} \nabla h_{\alpha^{(k)},\beta^{(k)}}(\bm s^{(k)})).
\end{align}
\end{definition}

The decoding process of the trainable discretized GF decoding is summarized in Algorithm \ref{GF_alg}.
Notably, every step in Algorithm 1 is differentiable, meaning that the automatic differentiation mechanism can be seamlessly applied to evaluate the gradients 
	of the trainable parameters.

Applying back-propagation with the Mean Squared Error (MSE) loss function enables the computation of gradients for the trainable parameters 
within the model. These gradients are essential to know how each parameter needs to be 
adjusted to minimize the loss. Gradient descent-based optimizers, such as Stochastic Gradient Descent (SGD) or Adam, 
leverage these gradients to update the trainable parameters. Incremental training is effective 
Details of how to train the deep unfolded model can be 
found in \cite{TISTA}.

\begin{figure}[htbp]
\begin{center}
\includegraphics[scale=0.6]{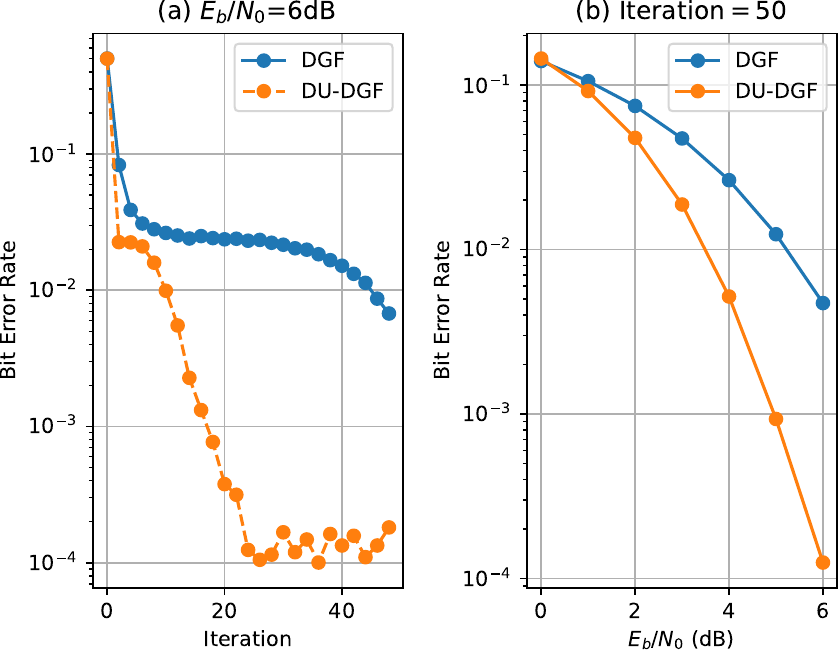}
\caption{BER performance of discretized GF decoding over AWGN channels. Left panel: BER as a function of the number of iterations. Right panel: BER as a function of $E_b/N_0$. }
\label{fig:du-demo}
\end{center}
\end{figure}

\begin{algorithm}[htbp]
 \caption{GF decoding with trainable parameters}
 \label{GF_alg}
 \begin{algorithmic}[1]
  \STATE $\bm s^{(0)} := \bm x_0$
  \FOR {$k := 0$ to $U-1$}
\STATE Set the following variables:
\begin{align} \nonumber
\bm z &:= \ln(\bm s^{(k)}), \\ \nonumber
\bm z_{-1} &:= \ln(\bm s^{(k)} - \bm 1), \\ \nonumber
\bm z_{+1} &:= \ln(\bm s^{(k)} + \bm 1), \\ \nonumber
\bm w &:= \ln (\bm H^T \left(\exp(2 \bm H \bm z)-\exp(\bm H \bm z) \right) ), \\ \nonumber
\bm g &:=\nabla L(\bm s^{(k)}; \bm y)
\end{align}
\STATE $\bm h := 4 \alpha^{(k)} \exp\left( \bm z_{-1} +  \bm z  +  \bm z_{+1} \right)+ 2 \beta^{(k)} \exp \left(\bm w - \bm z  \right)$
\STATE 	$\bm s^{(k+1)} := \bm s^{(k)} - \eta^{(k)} (\bm g + \gamma^{(k)} \bm h)$
  \ENDFOR
\STATE $\hat{\bm x} := \mbox{sign}(\bm{s}^{(U)})$
\STATE Output $\hat{\bm x}$
 \end{algorithmic} 
 \end{algorithm}

\section{GF decoding for MIMO channels}

\subsection{Generalized GF-ODE for MIMO channels}

In this section, 
we discuss an application of GF decoding for LDPC-coded massive MIMO channels.
Joint decoding of LDPC codes and MIMO channels 
is a practically important problem for future wireless systems such as beyond-5G/6G systems.

Let $\bm A \in \mathbb{R}^{N \times n}$ be a channel matrix.
Suppose that a received word $\bm{y} \in \mathbb{R}^N$ is given by
\begin{align}
	\bm{y} = \bm{A} \bm{x} + \bm{w},				
\end{align}
where $\bm{w} \in \mathbb{R}^N$ is a Gaussian noise vector,
the components of which follow an i.i.d. Gaussian distribution.
The channel input vector $\bm{x}$ is assumed to be a codeword of $C(\bm H)$,
which implies that we assume BPSK modulation.

In this problem setting, the PDF representing the channel is given by
\begin{align}
p(\bm y |\bm x) \equiv a \exp \left(- b \|\bm y - \bm A \bm x\|^2 \right),		
\end{align}
where $a$ and $b$ are real constants. Our objective function for the 
approximate MAP rule can be expressed as
\begin{align}
F_{\rm{MIMO}}(\bm x) \equiv \frac 1 2 \|\bm y - \bm A \bm x\|^2 +  \gamma h_{\alpha,\beta}(\bm x).	
\end{align}

\begin{corollary}
 $\nabla F_{\rm{MIMO}}(\bm x)$ is tensor-computable.
\end{corollary}
(Proof)
The gradient of the first term of $F_{\rm{MIMO}}$:
\begin{align}
\nabla L(\bm x; \bm y) = 
\nabla\frac 1 2 \|\bm y - \bm A \bm x\|^2 = \bm A^T (\bm A \bm{x} - \bm y)	
\end{align}
is tensor-computable. By Theorem 1, we immediately 
have the claim of the corollary. \hfill\fbox{}

The generalized GF-ODE for LDPC-coded MIMO channels thus becomes 
\begin{align}\label{MIMO_ODE}
	\frac{d\bm x}{dt} &= - (\bm A^T (\bm A \bm{x} - \bm y) + \gamma \nabla h_{\alpha,\beta}(\bm x)), 
\end{align}
where the initial condition is given by $\bm x(0) = \bm x_0$. In this case, 
the recursive formula of the discretized GF decoding is given by
\begin{align} \label{MIMOGD}
	\bm s^{(k+1)} &= \bm s^{(k)} - \eta (\bm A^T (\bm A \bm s^{(k)} - \bm y) + \gamma \nabla h_{\alpha,\beta}(\bm s^{(k)})),
\end{align}
where $\bm s^{(0)} = \bm x_0$.

\subsection{Experimental conditions}

In this subsection, we will explain the details of the 
numerical experiments for real-valued MIMO model.

Let $\bm A' \equiv \{a'_{i,j}\} \in \mathbb{C}^{\mu \times \nu}$ be
a channel matrix, where $a_{i,j}'$ is the fading coefficient corresponding to the path between the $j$th transmit antenna and the $i$th receive antenna.
We assume an i.i.d. model such that 
each component of $\bm A'$ follows a complex circular Gaussian distribution ${\cal CN}(0,1)$.
%
A QPSK modulation format is assumed for transmitted signals.
An {\em equivalent real-valued MIMO model} with BPSK modulation can be defined as
$\bm y = \bm A \bm x + \bm w$ where $\bm A$ is given by
$$
	\bm A \equiv
	\left[
	\begin{array}{cc}
		\mbox{Re}(\bm A') & -\mbox{Im}(\bm A') \\
		\mbox{Im}(\bm A') & \mbox{Re}(\bm A')
	\end{array}
	\right] \in \mathbb{R}^{\mu \times n},
$$
where $N = 2\mu$ and $n = 2\nu$.
The transmitted word $\bm x$ is randomly
chosen from $C(\bm H)$ according to the uniform distribution.
Each component of the Gaussian noise vector
$\bm w \in \mathbb{R}^N$ follows Gaussian distribution with zero mean and
variance $\sigma_w^2/2$. In this model, $\sigma_w^2$ is related to
the signal-to-noise ratio $\sf SNR$ by
$
	\sigma_w^2 = N/{\sf SNR}.	
$

In the following experiment, we used the regular (3,6)-LDPC code
 with $n =204$ and $m=102$.
The step-size parameter $\omega$ used in the gradient descent step
was set to
\begin{align}
\omega \equiv \frac{2}{\lambda_{min} + \lambda_{max}},  		
\end{align}
where $\lambda_{min}$ and $\lambda_{max}$ are the minimum and maximum
eigenvalues of $\bm A^T \bm A$, respectively.

For comparison, we exploited 
the MMSE detector defined as
\begin{align}
\hat {\bm{x}} \equiv \bm A^T (\bm A \bm A^T + (\sigma_w^2/2) \bm I)^{-1} \bm y.			
\end{align}
Furthermore, as a baseline for joint detection and decoding, 
we employed  the combination of the MMSE detector and 
BP decoding which is to be denoted by MMSE + BP.

\subsection{Numerical experiments}
In this subsection, we present experimental results on discretized GF (DGF) decoding for MIMO channels.

\begin{figure}[ht]
    \centering
    \includegraphics[width=0.48\textwidth]{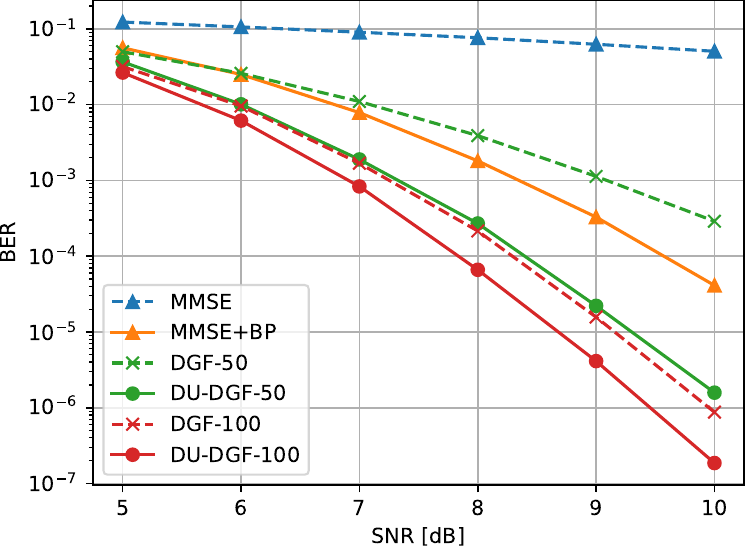}
    \caption{Bit error rate performances of DGF/DU-DGF decoding and baseline schemes $(n=204, m = 102)$. 
    }
    \label{fig:result-ber}
\end{figure}

Figure \ref{fig:result-ber} shows the bit error rate (BER) performances
of the proposed 'DGF' and its DU version, `DU-DGF', 
alongside baseline schemes.
The labels `DGF-50' and `DU-DGF-50' represent the BER performance evaluated at the $50$th iteration,
while `DGF-100' and `DU-DGF-100' represent the BER performance at the $100$th iteration.
The maximum number of iterations for BP decoding was set to $100$.

From the results, we can find that
the BER performance of the plain MMSE detector is far behind other methods that consider LDPC encoding (labeled with `MMSE+BP', `DGF', and `DU-DGF').
With a maximum number of iterations of $100$,
our proposed method (`DGF-100') achieves approximately a $1.6$dB advantage over `MMSE+BP' at BER$=10^{-4}$.
Additionally, with DU, the adjustment of the step size for each iteration can speed up the convergence of DGF,
further widening the BER performance gap between our proposed method and the `MMSE+BP' approach.
Meanwhile, by comparing the results at the $50$th and $100$th iterations,
we also found that the improvement of `DU-DGF' is more obvious in fewer iterations.

%
%


\section{Score-Based Channel Learning}

\subsection{Overview}

A trajectory of the search vector of GF decoding is determined by the vector field created 
by $-\nabla \ln p(\bm y | \bm x)$ and $\gamma \nabla h_{\alpha,\beta}(\bm x)$. 
Essentially, these vector fields dictate the dynamics of the GF decoding process.
If the negative log-likelihood function of the channel, $\nabla \ln p(\bm y | \bm x)$, is perfectly known and can be computed using tensor-based computation, then implementing GF decoding becomes relatively simple.
	For example, channels with correlated Gaussian noise and certain nonlinear vector channel 
	fall under this class, as $\nabla \ln p(\bm y | \bm x)$ for these types of channels can be easily expressed in a simple formula \cite{wadayama23b}.	Conversely, if the negative log-likelihood function for the channel is unknown, it is necessary to learn this function from data. The section discusses {\em channel learning} for GF decoding from data on inputs and outputs of the target channel.

For a given probability density function $p(\bm x)$, 
\begin{align}
s(\bm x) \equiv \nabla \ln p(\bm x)
\end{align}
 is referred to as the {\em score function} of $p$ \cite{Song}\cite{Song2021}. It is known that by combining the score function $s(\bm x)$ with Langevin sampling, it is possible to efficiently generate random vectors that follow $p(\bm x)$. 
 Generative models based on score functions 
 is currently known as {\em score-based diffusion models}.
In the context of generative models based on score functions, a neural network model approximating $s(\bm x)$ 
is learned from given data. It is known that modeling the score function with a neural network \cite{Song} offers more benefits than directly modeling the probability density function $p(\bm x)$ itself with a neural network.

If one wishes to use GF decoding in situations where the true likelihood function $p(\bm y|\bm x)$ of the channel 
is unknown, it is anticipated that 
learning the {\em conditional score function}
\begin{align}
s(\bm x; \bm y) \equiv	\nabla_{\bm x}\ln p(\bm y|\bm x)	
\end{align}
from data is beneficial. After leaning $s(\bm x; \bm y)$,
the model $\hat{s}_\theta(\bm x; \bm y)$ can be used as 
\begin{align}
\nabla L(\bm x; \bm y) \equiv -\hat{s}_\theta(\bm x; \bm y)		
\end{align}
in the GF decoding process. Research on learning techniques for score functions, such as score matching methods \cite{Song2021}, is actively being pursued in the context of diffusion models, and it is believed that these studies could enable the realization of novel channel learning techniques suitable for GF decoding.

\subsection{Channel score model}
\subsubsection{Requirements}
Modeling the conditional score function $s(\bm x; \bm y)$ with a neural network is an intuitive strategy. 
The model must meet certain criteria:
\begin{itemize}
\item  It should be tensor-computable.
\item A shallow network architecture is favored to ensure efficient inference processes.
\item To maintain training efficiency, the model should be designed with a limited number of parameters.
\item  If prior knowledge about the statistical characteristics of the channel exists, the model architecture should reflect this information. For instance, exploiting the memoryless property could greatly simplify the model architecture.
\item There should be an effective method for training the model.
\end{itemize}

\subsubsection{Neural network model}

Although a variety of neural network models could fulfill the specified requirements described above, 
for the sake of simplicity, our discussion will focus on employing a simple feedforward neural network.
Note that, depending on the characteristics of channels, other models might be more appropriate for channel learning tasks.

In the following discussion, we assume an additive channel where 
there exists a PDF $Q$ satisfying 
\begin{align}
	p(\bm y| \bm x) = Q(\bm x - \bm y)
\end{align}
for the sake of simplicity of the argument, i.e.,
the received word $\bm y$ is given by 
$
	\bm y = \bm x + \bm e,	
$
where $\bm e \sim \bm Q(\bm e)$.
We further assume a {\em 
segment-wise independence and an identically distributed (i.i.d.)
memoryless property} characterized by the equation:
\begin{align}
	p(\bm y|\bm x) = \prod_{i=1}^K p(\bm y_i'|\bm x_i')= \prod_{i=1}^K q(\bm x_i' - \bm y_i'),
\end{align}
where $\bm x \in \mathbb{R}^n$ and $\bm y \in \mathbb{R}^N$ ($n = N$) are segemented into:
\begin{align}
\bm x \equiv (\bm x_1', \bm x_2',\ldots, \bm x_K'),\quad
\bm y \equiv (\bm y_1', \bm y_2',\ldots, \bm y_K'),
\end{align}
where $q: \mathbb{R}^\nu \rightarrow \mathbb{R}$ is a PDF ($n = \nu K$).
In the discussions that follow, we will use boldface letters with a prime symbol to denote vectors 
that correspond to the segment.

Given these assumptions, the conditional score function can be broken down as follows:
\begin{align}
s(\bm x; \bm y) &= \nabla_{\bm x}\ln p(\bm y|\bm x) \\
&= \nabla_{\bm x}\sum_{i=1}^N\ln p(\bm y_i'|\bm x_i') \\
&= \nabla_{\bm x}\sum_{i=1}^N\ln q(\bm x_i' - \bm y_i') \\
&= 
\left(
\begin{array}{c}
\nabla_{\bm x_1'}\ln q(\bm x_1' - \bm y_1') \\
\vdots \\
\nabla_{\bm x_K'}\ln q(\bm x_K' - \bm y_K') \\
\end{array}
\right) \\
&= 
\left(
\begin{array}{c}
\varsigma(\bm x_1' - \bm y_1') \\
\vdots \\
\varsigma(\bm x_K' - \bm y_K') \\
\end{array}
\right),
\end{align}
where 
\begin{align}
\varsigma(\bm e')  \equiv \nabla_{\bm e'}\ln q(\bm e')	
\end{align}
because
\begin{align}
\nabla_{\bm x_i'}\ln q(\bm x_i' - \bm y_i')	= \nabla_{\bm e_i'}\ln q(\bm e_i')	
\end{align}
holds when $\bm e'_i = \bm x_i' - \bm y_i'$.
The function $\varsigma(\bm e')$ is referred to as 
the {\em segmented score function}.

Next, we delve into utilizing a neural network to approximate the segmented score function $\varsigma$. We define a feedforward neural network model $\hat{\varsigma}_\theta: \mathbb{R}^\nu  \rightarrow \mathbb{R}^\nu$ as follows:
\begin{align} 
 \hat{\varsigma}_\theta(\bm e') &=  \bm h_L, \\
 \bm h_{i+1} &= g_i(\bm W_i \bm h_i + \bm b_i),\quad i = 1,2,\ldots, L-1, \\ 
 \bm h_1 &= \bm e',
\end{align}	
where each $g_i$ represents an activation function, and the set $\theta$ encompasses all trainable parameters in the model. The $\bm W_i$ matrices and $\bm b_i$ vectors are the trainable weights and biases, respectively. 
The aim is for the {\em segmented score model} $\hat{\varsigma}_\theta$ to be trained so that
\begin{align}
s(\bm x; \bm y) \approx 
\left(
\begin{array}{c}
\hat \varsigma_\theta(\bm x_1' -  \bm y_1') \\
\vdots \\
\hat \varsigma_\theta(\bm x_K' - \bm y_K') \\
\end{array}
\right),
\end{align}
effectively modeling the segmented score function for each segment.


\subsection{Score matching learning}

We first briefly outline the idea of the score matching learning according to \cite{Song}\cite{Song2021}. 
Assume that we wish to learn the score function of $p(\bm x)$ from the data following $p(\bm x)$.
It is natural to minimize the expected loss function:
\begin{align}
	\frac 1 2\mathbb{E}_{\bm x \sim p(\bm x)}[\|s_\theta(\bm x) -  \nabla \log p(\bm x)\|^2_2]
\end{align}
for tuning the trainable parameter in the model $s_\theta(\bm x)$ but evaluation of $\nabla \log p(\bm x)$ is impossible as $p(\bm x)$ is unknown. The above loss function is known to be equivalent to the following loss function up to a constant:
\begin{align}
	\frac 1 2\mathbb{E}_{\bm x \sim p(\bm x)}\left[\mbox{tr}\left(\nabla s_\theta(\bm x)\right) + \frac{1}{2} \|s_\theta(\bm x) \|^2_2 \right].
\end{align}
In principle, the above expectation can be evaluated but is not suitable 
for learning relatively large model such as deep neural network. In this paper,
we use the {\em denoising score matching loss} \cite{Song}  
defined by
\begin{align}
\frac 1 2\mathbb{E}_{\bm x \sim p(\bm x)}	\mathbb{E}_{\tilde{\bm x} \sim {\cal N}(\bm x, \sigma^2 \bm I)}
\left[\left\|\bm s_\theta(\tilde{\bm x}) + \frac{\tilde{\bm x} - \bm x}{\sigma^2} \right\|_2^2 \right].
\end{align}
It is anticipated that the trained model can precisely approximates 
the true score function if the noise variance $\sigma^2$
is small enough \cite{Song}.

\subsection{Details of score-based channel learning}

Assume that we have a mini-batch generator producing 
a random mini-batch for additive error batch:
$
	{\cal E} \equiv (\bm e_1', \bm e_2',\ldots, \bm e_D'), 
$
where each error vector $\bm e_i'$ is sampled according to the PDF $q$.
This mini-batch generator could be designed based on a theoretical probabilistic channel model or using actual data collected from the real target channel.

The details of score-based channel learning is summarized in Algorithm \ref{score_matching},
which is based on the standard method for denoising socre matching \cite{Song}.
To update trainable parameters, we can utilize standard optimization techniques like stochastic gradient descent (SGD) and Adam. The noise variance $\sigma^2$ specifying the disturbing noise level is a hyperparameter that needs to be determined before the training begins.
\begin{algorithm}[htbp]
 \caption{Score-based channel learning}
 \label{score_matching}
 \begin{algorithmic}[1]
  \STATE Initialize the trainable parameters in $\theta$.
  \FOR {$k := 1$ to $U$}
\STATE Generate a random mini-batch ${\cal E}$.
\STATE Generate disturbed samples:
\begin{align}
	\tilde{\bm e}_d' = {\bm e}_d' + \bm n_d
\end{align}
for $d \in [D]$ where $\bm n_d \sim {\cal N}(0, \sigma^2 \bm I)$.
\STATE 	Evaluate the score matching loss:
\begin{align}
	{\cal L}(\theta; {\cal X}, {\cal Y}) \equiv \frac{1}{D}\sum_{d = 1}^D\left\|\hat{\varsigma}_\theta(\tilde{\bm e}_d') + \frac{\tilde{\bm e}_d' - \bm e_d'}{\sigma^2} \right\|_2^2.
\end{align}
	\STATE Perform  back-propagation to evaluate the gradient of ${\cal L}$ 
	respect to the trainable parameters in $\theta$.
	\STATE Update the the trainable parameters in $\theta$.
  \ENDFOR
 \end{algorithmic} 
 \end{algorithm}
 
As a result of the training process, we acquire a trained model 
denoted by $\hat{\varsigma}_{\theta^*}$. The gradient of
negative log likelihood function $L$ can then be approximated using this trained model as: 
\begin{align}
\nabla L(\bm x; \bm y) \approx
\left(
\begin{array}{c}
-\hat \varsigma_{\theta^*}(\bm x_1' - \bm y_1') \\
\vdots \\
-\hat \varsigma_{\theta^*}(\bm x_K' -  \bm y_K') \\
\end{array}
\right).
\end{align} 
This approximated negative log likelihood function can be used in the GF decoding process.

There are a couple of strategies to training the whole GF decoding process:
\begin{enumerate}
\item {\em Serial training}: The most straightforward method involves initially training the model $\hat \varsigma_{\theta}$, and then using deep unfolding to train the trainable parameters specified in (\ref{GFGD_trainable}), while keeping $\hat \varsigma_{\theta}$ fixed.
\item {\em Fine tuning}: Another approach starts with training the model $\hat \varsigma_{\theta}$ as well, but then combines deep unfolding training to optimize both the set of trainable parameters in (\ref{GFGD_trainable}) and the trainable parameters within $\hat \varsigma_{\theta}$.
\end{enumerate}

It is crucial to emphasize that, in the training process outlined in Algorithm \ref{score_matching}, the disturbed samples are generated using a single noise level. However, the literature on score-based diffusion models \cite{Song}\cite{Song2021} has established that a gradual change in the noise level is essential for accurately approximating a complex score function. Consequently, further development of Algorithm \ref{score_matching} to incorporate varying noise levels constitutes a significant area for future research, as it has the potential to enhance the accuracy and effectiveness of the score-based channel learning.

\subsection{Numerical example}

\begin{figure}[ht]
    \centering
    \includegraphics[width=0.48\textwidth]{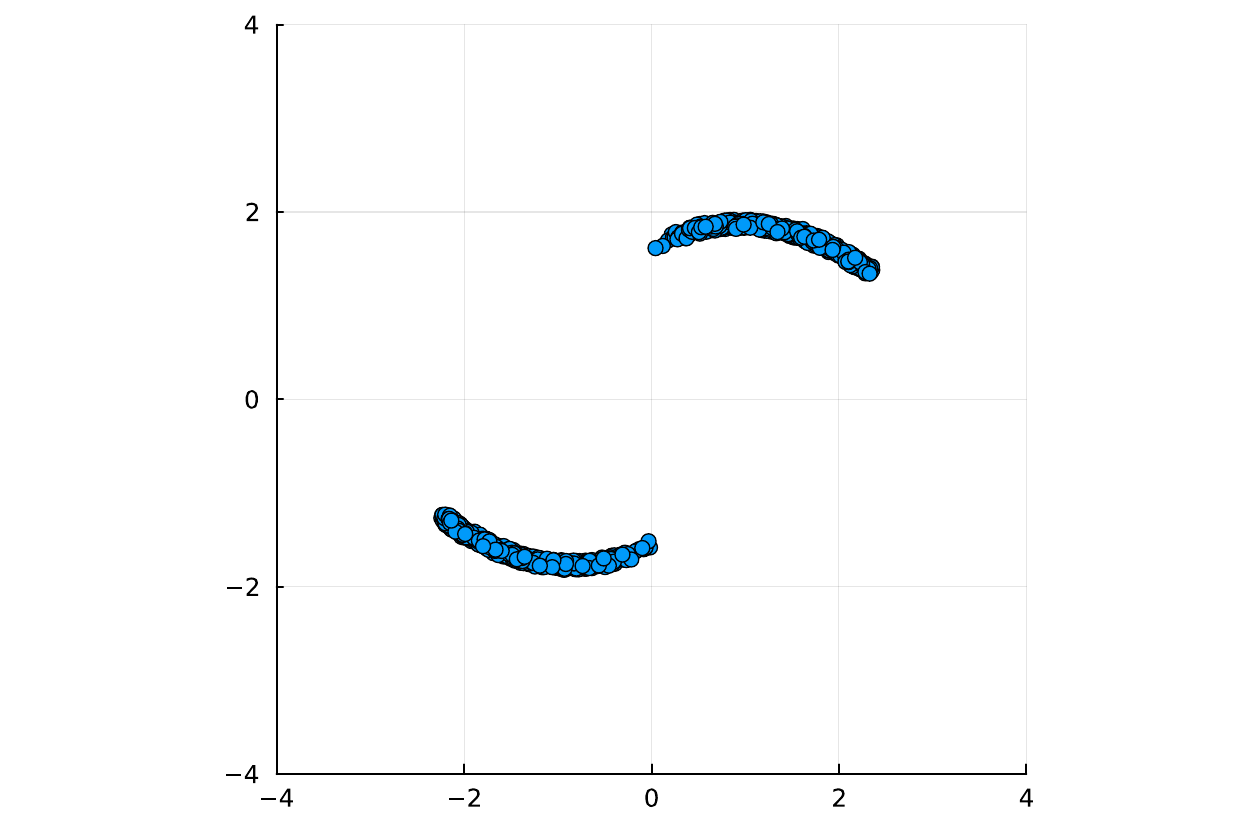}
    \caption{Correlated error candidates (1000 2D-points). One of the 2D-points is randomly realized in our channel model.}
    \label{fig:orgdata}
\end{figure}

In this subsection, we explore a numerical example for GF decoding utilizing the trainable score-based gradient. We focus on a scenario involving artificial two-dimensional correlated noises, i.e., the segment size is $\nu = 2$. A total of 1000 two-dimensional error candidates, $\bm e_1', \ldots, \bm e_{1000}'$, are illustrated in Fig.~\ref{fig:orgdata}. The segment-wise channel model is defined as
\begin{align}
	\bm y' = \bm x' + \bm e',
\end{align}
where $\bm e'$ is selected uniformly at random from the error candidates.
In this example, we utilize a serial training approach 
and opt not to employ deep unfolding for additional optimization.

Initially, we train the model $\hat \varsigma_{\theta^*}(\bm e')$ using mini-batches generated randomly in accordance with the probabilistic channel model previously defined. Our model architecture comprises a three-layer ($L=3$) feedforward neural network equipped with the ReLU activation function, and we configured each layer of the network to have 64 hidden units. During the training phase, we set the mini-batch size to $100$ and carried out 10,000 training iterations. For the adjustment of trainable parameters, the Adam optimizer with learning rate $0.005$ was employed. For generating the disturbed samples,
	Gaussian noises with $\sigma = 0.3$ were used in Algorithm \ref{score_matching}.

\begin{figure}[ht]
    \centering
    \includegraphics[width=0.48\textwidth]{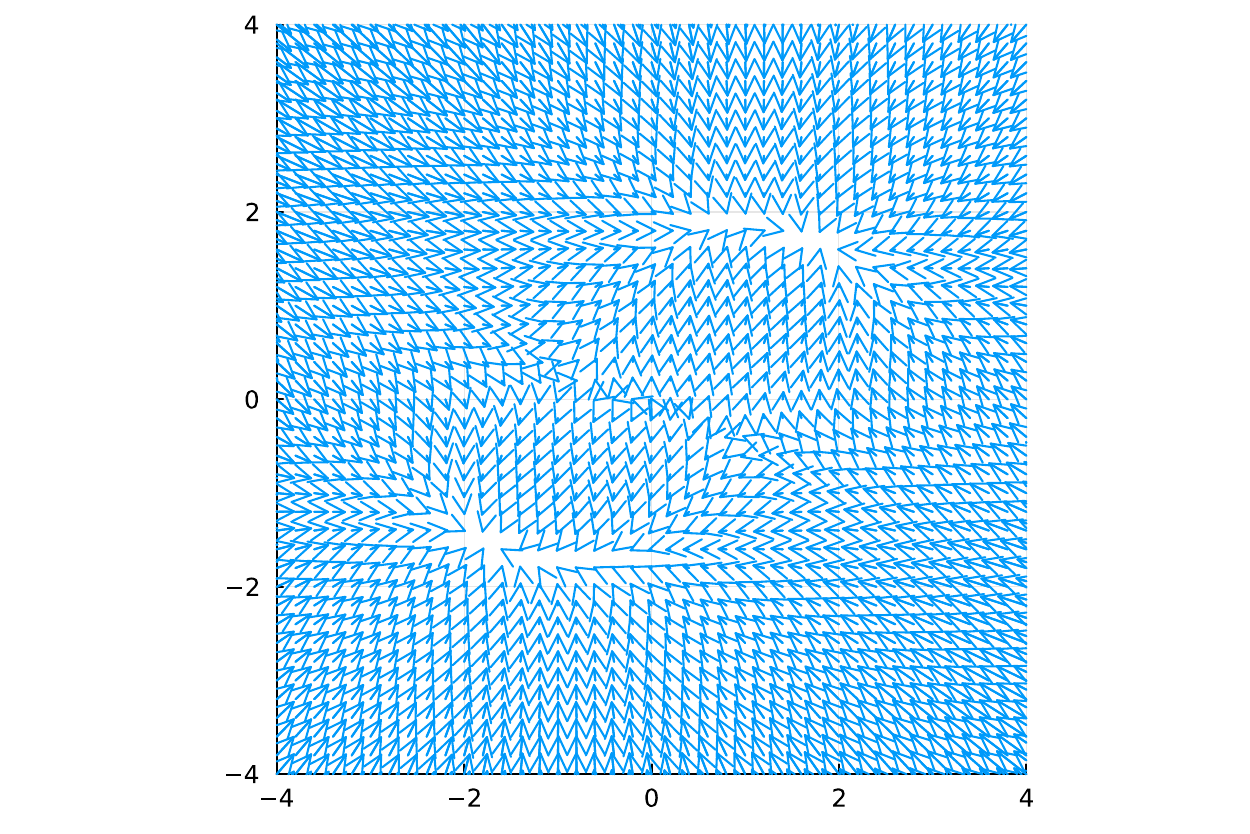}
    \caption{Learned vector field given by $\hat \varsigma_{\theta^*}$. }
    \label{fig:vfield}
\end{figure}

Figure \ref{fig:orgdata} presents the vector field generated by the trained model $\hat \varsigma_{\theta^*}$. Upon comparing Figs \ref{fig:orgdata} and \ref{fig:vfield}, it becomes apparent that regions near the error candidate points correspond to attractors in the learned vector field. 
Specifically, a vector in Fig.~\ref{fig:orgdata} points towards the nearest error candidate point
which is consistent with the ascending direction of the true 
segmented score function. This indicates that the trained model $\hat \varsigma_{\theta^*}(\bm e')$ effectively encapsulates the characteristics of the segmented score function $\varsigma(\bm e')$.

We conducted experiments on the discretized GF decoding process using the trained model $\hat \varsigma_{\theta^*}(\bm e')$. For these experiments, a $(3,6)$-regular LDPC code with $n = 204$ and $m = 102$ was utilized. The configuration for the GF decoder was set as follows: $\alpha = \beta = 1, \eta = 0.05$ ,and $\gamma = 1.0$ were chosen. The initial state of the GF decoder was initialized by a random vector $\bm x_0$ drawn from a normal distribution ${\cal N}(\bm 0, 0.1^2 \bm I)$.

\begin{figure}[ht]
    \centering
    \includegraphics[width=0.48\textwidth]{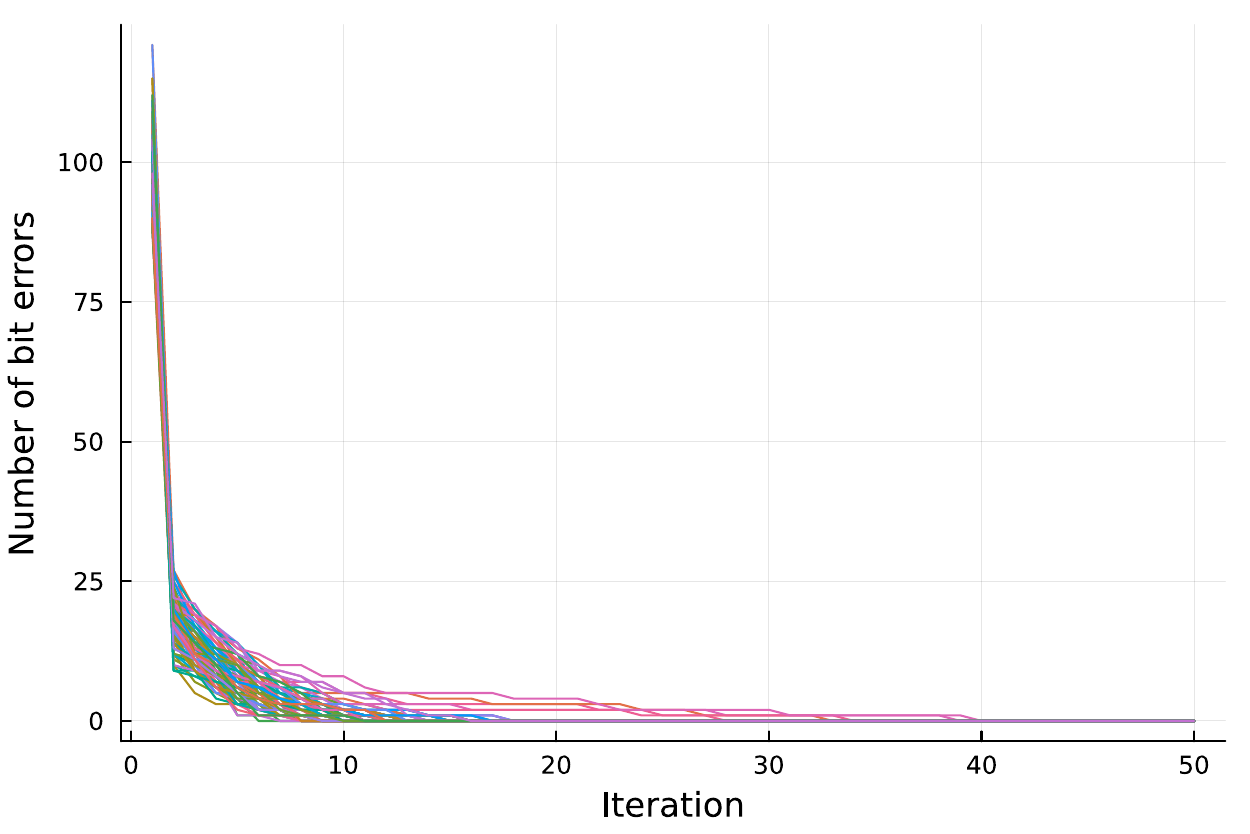}
    \caption{Number of bit errors as a function of the number of iterations (100 trials, (3,6)-regular LDPC code with $n = 204, m = 102$). }
    \label{fig:BitError}
\end{figure}

Figure \ref{fig:BitError} displays the discrepancy count between
$
\hat{\bm x}^{(k)} \equiv \mbox{sign}(\bm{s}^{(k)})
$
and the transmitted word $\bm x$, i.e., the estimated number of errors as a function of the iteration count $k$. In this experiment, we conducted 100 GF decoding trials, and each curve in Fig.~\ref{fig:BitError} illustrates the change in the number of bit errors for each trial. It is noticeable that all curves exhibit a sharp decrease during the initial few iterations, followed by a more gradual reduction. This indicates that the gradient descent processes are effectively functioning with the learned score function. In all trials, errors were eliminated before reaching 50 iterations.

\subsection{Segmented linear vector channel}

We here briefly discuss the case where the segment sizes are not equal, i.e., $N \ne  n$. 
An important channel in this class is a segmented linear vector channel defined by 
\begin{align}
	\bm y' = \bm A \bm x' + \bm e',
\end{align}
where $\bm A \in \mathbb{R}^{N \times n}$ and $\bm e' \sim q(\bm e')$.
For example, a MIMO channel with correlated additive noises are included in this class.
In this case, it is expected that the conditional score function $s(\bm x; \bm y)$ can be well 
approximated by 
\begin{align}
s(\bm x; \bm y) \approx 
\left(
\begin{array}{c}
\bm A^{T} \hat \varsigma_\theta(\bm A \bm x_1' -  \bm y_1') \\
\bm A^{T} \hat \varsigma_\theta(\bm A \bm x_2' - \bm y_2') \\
\vdots \\
\bm A^{T} \hat \varsigma_\theta(\bm A \bm x_K' - \bm y_K') \\
\end{array}
\right),
\end{align}
where $\hat \varsigma_\theta(\bm e')$ is a neural network model approximating 
the segmented score function 
$\varsigma(\bm e') = \nabla_{\bm e'}\ln q(\bm e')$.


This framework is capable of handling multi-valued signal constellations, such as Quadrature Amplitude Modulation (QAM). A straightforward example is as follows:
Consider $\bm x' = (x_1, x_2, x_3, x_4)$ and $\bm y' = (y_1, y_2)$ where
\begin{align}
	y_1 = 2x_1 + x_2,\quad	y_2 = 2x_3 + x_4.
\end{align}
Given the definition, it is evident that $y_i(i = 1,2)$ takes a value in $\{-3, -1, 1, 3\}$ because $x_i \in \{1, -1\} (i=1,2)$. Thus, a 16 QAM signal constellation can be obtained through the linear mapping
$
	\bm y' = \bm A \bm x',
$
with $\bm A$ defined as
\begin{align}
	\bm A \equiv 
	\left(
	\begin{array}{cccc}
		2 & 1 & 0 & 0 \\
		0 & 0 & 2 & 1 \\
	\end{array}
	\right).
\end{align}
This approach is expected to be applicable to numerous practical wireless communication systems.

%
%
%
%

\section{Conclusion}

This paper introduces the GF decoding algorithm, which has a tensor-friendly design, making it particularly well-suited for implementation on next-generation AI accelerators. The underlying principle of GF decoding is remarkably universal, allowing for its application to a wide range of non-trivial channels, provided that the gradient of the negative log-likelihood function can be efficiently evaluated. Another significant advantage of our proposed method lies in its suitability for deep unfolding. This is facilitated by the ease of evaluating the gradient of the internal trainable parameters, which can be efficiently handled using automatic differentiation mechanisms or back-propagation, as every component of the algorithm is differentiable.

In massive MIMO channels encoded with an LDPC code, the proposed algorithm 
demonstrates competitiveness with established detection methods, 
such as MMSE + BP. When considering the balance between decoding complexity 
and decoding performance, 
GF decoding shows considerable promise for LDPC-coded MIMO channels.
Traditionally, signal processing algorithms in a receiver have been designed following a {modular design} principle. Typically, components like a MIMO signal detector and an LDPC decoder are designed separately and then integrated. However, the principle of GF decoding facilitates a {\em codesign} principle \cite{Worthen01}, potentially enhancing the overall performance of the receiver. The methodology outlined in this paper may pave the way for innovative strategies in developing jointly designed signal processing algorithms within a receiver.

Score-based channel learning offers a novel approach to learn the probabilistic nature of the target channel. A neural network can be employed to approximate a conditional score function, and the resulting model can be directly utilized in GF decoding as the gradient of the negative log-likelihood function. As the importance of handling unknown channels grows, data-driven methods for channel learning will become increasingly crucial in the field of communications.

Error correcting codes and their decoding methods have evolved in tandem with the advancements in foundational hardware technology. In the era when only small-scale integrated circuits were available, algebraic codes were highly valued for their compatibility with the hardware limitations of the time. As technology progressed to enable the assembly of large-scale ASICs, the implementation of decoders for LDPC codes in ASICs became feasible, ushering in a golden era for LDPC codes. Looking ahead, the hardware suited for AI learning and inference processes, which are poised to be ubiquitous in the near future, should not only be capable of supporting the training and inference processes of generative AI models but also be versatile enough to accommodate a wide range of signal processing tasks. As AI hardware continues to evolve, error correcting codes and their decoding methods are expected to follow suit in the 21st century, adapting to the prevailing hardware environments of their times.
\section*{Acknowledgement}
This work was supported by JSPS KAKENHI Grant-in-Aid for Scientific Research (A) Grant Number JP22H00514. 
The authors appreciate the inspiring discussions on the initial version of GF decoding with Dr. Ayano Nakai-Kasai and Mr. Kensho Nakajima.

\section*{Appendix}

\subsection*{Tensor-computable BP decoding}

It is possible to formulate log-domain BP decoding 
in tensor-computable form.
Let the set of edges contained in the Tanner graph be defined as
\begin{align}
	E \equiv \{ (i,j) \in [m] \times [n] \mid H_{ij} = 1 \}.
\end{align}
In the following, the total number of edges contained in the Tanner graph is denoted by \(e = |E|\). The number of edges \(e\) is equal to the number of 1's contained in \(\bm H\).
Now, assume that a bijection
\(\phi: E \rightarrow [e]\) is given. That is, the edges contained in \(E\) are numbered by \(\phi\).
Furthermore, for all \(k \in [e]\), let \(p_k, q_k\) be defined as the quantities that satisfy \(\phi((p_k, q_k)) = k\).

Define the matrix \(\bm U = \{U_{jk}\}\in \{0,1\}^{n \times e}\) and the matrix \(\bm V = \{V_{ik}\}\in \{0,1\}^{m \times e}\) as
\begin{align}
	U_{jk} &\equiv 
	\begin{cases}
	1, & \text{if } q_k = j	 \\
	0, & \text{otherwise,} \\
	\end{cases}
\end{align}

\begin{align}
	V_{ik} &\equiv 
	\begin{cases}
	1, & \text{if } p_k = i	 \\
	0, & \text{otherwise}. \\
	\end{cases}
\end{align}

As an example, define the parity-check matrix \(\bm H\) as
\begin{align}
\bm H \equiv \left[
\begin{array}{cccccc}
1 & 1 & 1 & 0 & 0 & 0 \\
0 & 0 & 1 & 1 & 0 & 0 \\
0 & 0 & 0 & 1 & 1 & 1 \\
\end{array}
\right].
\end{align}
Define the numbering of edges in the Tanner graph as
\begin{align} \nonumber
  \phi(1, 1) = 1,
  \phi(1, 2) = 2,
  \phi(1, 3) = 3,
  \phi(2, 3) = 4, \\ \label{phi_edge}
  \phi(2, 4) = 5,
  \phi(3, 4) = 6,
  \phi(3, 5) = 7,
  \phi(3, 6) = 8.
\end{align}
In this case, the matrices $\bm U$ and $\bm V$ are 
given by:
\begin{align}
\bm U = \left[
\begin{array}{cccccccc}
1 & 0 & 0 & 0 & 0 & 0 & 0 & 0 \\
0 & 1 & 0 & 0 & 0 & 0 & 0 & 0 \\
0 & 0 & 1 & 1 & 0 & 0 & 0 & 0 \\
0 & 0 & 0 & 0 & 1 & 1 & 0 & 0 \\
0 & 0 & 0 & 0 & 0 & 0 & 1 & 0 \\
0 & 0 & 0 & 0 & 0 & 0 & 0 & 1 \\
\end{array}
\right]
\end{align}
and
\begin{align}
\bm V = \left[
\begin{array}{cccccccc}
1 & 1 & 1 & 0 & 0 & 0 & 0 & 0 \\
0 & 0 & 0 & 1 & 1 & 0 & 0 & 0 \\
0 & 0 & 0 & 0 & 0 & 1 & 1 & 1 \\
\end{array}
\right].
\end{align}

These matrices $\bm{U}$ and $\bm{V}$ are key matrices in rewriting the belief propagation decoding into a tensor-computable form.
The details of tensor-friendly BP decoding are summarized in Algorithm~\ref{diff_BP}. The vectors $\bm{\alpha}$ and $\bm{\beta}$ represent check-to-variable and variable-to-check messages, respectively. The vector $\bm{\lambda}$ denotes the log likelihood ratio vector. We can see that a dominant computation of the BP decoding process is a matrix-vector product of $\bm{U}^T \bm{U} - \bm{I} \in \mathbb{R}^{e \times e}$ and $\bm{\alpha} \in \mathbb{R}^e$.

Assume that 
\begin{itemize}
\item Batch processing with the batch size $D$.
\item In a GPGPU, a single thread handles dot product computations of $O(e)$ as in the thread computation model in Subsection~\ref{thread_model}.
\end{itemize}
In this case, we need $eD$ threads for executing a BP decoding process. Recall that discretized GF decoding requires only $mD$ threads and that each thread handles $O(n)$ processing. Because $e > n > m$, the tensor-computable BP decoding in Algorithm~\ref{diff_BP} is more resource-demanding than discretized GF decoding.

\begin{algorithm}
 \caption{Tensor-compuatable BP Decoding}
 \label{diff_BP}
 \begin{algorithmic}[1]
  \STATE \textbf{Initialization:} Set $\bm \alpha \equiv \bm 0, \bm \beta \equiv \bm 0$
  \FOR {$i = 1$ to $L$}
  	\STATE \textbf{Variable Node Processing:}
  	\begin{align}
	\bm \beta \equiv (\bm U^T \bm U - \bm I) \bm \alpha + \bm U^T \bm \lambda
\end{align}
	\STATE \textbf{Check Node Processing:}
\begin{align}
	\bm \alpha_{\sf abs} &\equiv 2 \tanh^{-1}\left\{
		\exp \left(
			(\bm V^T\bm V - \bm I) 
				\log \left|
					\tanh \frac{\bm \beta}{2}
					\right|
				\right)
		\right\} \\
	\bm \alpha_{\sf sign} &\equiv 
		\left\{
		\bm 1 - 2\bm V^T
		\mbox{mod}
				\left(
				\bm V(\bm 1 - \mbox{sign}\bm \beta)/2
				\right)
		\right\}\mbox{sign}\bm \beta \\
		\bm \alpha &\equiv \bm \alpha_{\sf sign} \odot \bm \alpha_{\sf abs}
\end{align} 
The function \(\mbox{bmod}\) represents the real remainder modulo 2, which is defined by
\begin{align}
\mbox{bmod}(x) \equiv x-2 \left\lfloor \frac{x}{2} \right\rfloor.
\end{align}

  \ENDFOR
  \STATE \textbf{Calculation of Posterior Probability Ratio:}
  \begin{align}
	\bm \gamma \equiv \bm U \bm \alpha + \bm \lambda
\end{align}
 \end{algorithmic} 
 \end{algorithm}


\begin{thebibliography}{00}
    \bibitem{wadayama23}T. Wadayama, K. Nakajima, and A. Nakai-Kasai, ``Gradient flow decoding for LDPC codes,'' 2023 International Symposium on Topics in Coding (ISTC2023), Brest, France, 2023. 

    \bibitem{wadayama24}T. Wadayama and L. Wei, ``Generalized gradient flow decoding and and Its Tensor-Computability,'' International Symposium on Information Theory (ISIT2024), Athens, 2024. 

	\bibitem{Gallager63}
  R. G. Gallager,
  ``Low density parity check codes,''
	MIT Press, 1963.
    \bibitem{Peltonen} E. Peltonen et al. , ``6G white paper on edge intelligence,’’ 6G Research Visions, No. 8. University of Oulu,  2020.
	\bibitem{OpenAI23}
	OpenAI, ``GPT-4 technical report, '' ArXiv:2303.08774v4, 2023.
  \bibitem{Reuther22}
  A. Reuther, P. Michaleas, M. Jones, V. Gadepally, S. Samsi, J. Kepner,
   ``AI and ML accelerator survey and trends,''
   IEEE High Performance Extreme Computing Conference (HPEC), 2022.
    \bibitem{opt}J. Carolan et al.,``Universal linear optics,''
    Science, vol.349, no. 6249, pp.711-716, 2015.
    \bibitem{uni} J. Capmany, D. Pérez, Programmable Integrated Photonics. Oxford University Press, 2020.
    \bibitem{mzi1}X. Pengfei and Z. Zhou,
    ``Silicon-based optoelectronics for general-purpose matrix computation: a review,'' Advanced Photonics, vol.4, no. 4, pp.044001-044001, 2022.
   \bibitem{Zhang2021}
  	H.~Zhang and  M.~Gu and  X.~D.~Jiang and et al.,
   ``An optical neural chip for implementing complex-valued neural network,''
   Nature Commun., vol.12, 2021.
   
   \bibitem{Prabhu20}
    M.Prabhu et al., 
    ``Accelerating recurrent Ising machines in photonic integrated circuits,''
    Optica, no.7, pp.551-558, 2020.
    
    \bibitem{Wang2023}
{S.~Wang, Y.~Luo, P.~Zuo,  L.~Pan, Y.~Li, and Z.~Sun,
``In-memory analog solution of compressed sensing 
recovery in one step,'' 
Science Advances, vol.9, no.50, 2023.}

\bibitem{Zuo2023}
{P.~Zuo, Z.~Sun, R.~Huang,
``Extremely-fast, energy-efficient massive MIMO precoding with analog RRAM matrix computing,''
IEEE Transactions on Circuits and Systems II: Express Briefs,
vol.7, no.7, pp.2335-2339, 2023.}

    \bibitem{wadayama10}T. Wadayama, K. Nakamura, M. Yagita, Y. Funahashi, S. Usami, I. Takumi, ``Gradient descent bit flipping algorithms for decoding LDPC codes'', IEEE Trans. Comm., pp.1610-1614, vol.58, no.6, June (2010)
	\bibitem{wadayama23b}
    T. Wadayama and S. Takabe, ``Proximal decoding for LDPC codes,'' IEICE Transactions on Fundamentals of Electronics, Communications and Computer Sciences, vol. E106-A, no. 3 pp. 359-367 (2023). 

   	\bibitem{Feldman03}
      J. Feldman, ``Decoding error-correcting codes via linear programming,''
 		Massachusetts Institute of Technology, Ph. D. thesis, 2003.
 	 \bibitem{Vontobel08}
     P. O. Vontobel,
     ``Interior-point algorithms for linear-programming decoding,''
     IEEE Information Theory and Applications Workshop, 2008.
     \bibitem{Wadayama10b}
     T.Wadayama, 
   	 ``Interior point decoding for linear vector channels based on convex optimization,''
  	IEEE Transactions on Information Theory,
  	vol.56, no.10, pp.4905--4921, 2010.
	\bibitem{Sundararajan14}
  	G. Sundararajan and C. Winstead and E. Boutillon,
  	``Noisy gradient descent bit-flip decoding for {LDPC} codes,''
  	IEEE Transactions on Communications,
  	vol.62, no.10, pp.3385--3400, 2014.
    \bibitem{Zhang13}
  	X. Zhang and P. H. Siegel, 
    ``Efficient iterative {LP} decoding of {LDPC} Codes with alternating direction method of multipliers,''
    IEEE International Symposium on Information Theory (ISIT), 2013.

	\bibitem{Barman13}
  S. Barman, X. Liu, and S. C. Draper, and B. Recht,
  ``Decomposition methods for large scale LP decoding,''
  IEEE Transactions on Information Theory,
  vol.59, no.12, pp.7870-7886, 2013.

	\bibitem{Jiang16}
	Y. Jiang, P. T. S. DeVore, A. Mahjoubfar, and B. Jalali, 
	``Analog logarithmic computing primitives with silicon photonics,''
	 in Conference on Lasers and Electro-Optics, OSA Technical Digest, 2016.
  	 \bibitem{Worthen01}
    A.P. Worthen and W.E. Stark,  ``Unified design of iterative receivers using factor graphs,''
  IEEE Transactions on Information Theory, vol.47, pp.843--849, 2001.
  
  \bibitem{NumericalMethods}
D.~F.~Griffiths, and D.~J.~Higham,
``Numerical methods for ordinary differential equations, '' Springer, 2010. 

\bibitem{MacKay}
D.~J.~C. MacKay, ``Encyclopedia of sparse graph codes [online],''
  \emph{Available: http://www.inference.phy.cam.ac.uk/mackay/codes/data.html}.

\bibitem{LISTA} K.~Gregor and Y.~LeCun,
``Learning fast approximations of sparse coding,''
 \textit{Proc. 27th Int. Conf. Machine Learning}, pp. 399--406, 2010.


\bibitem{9020494}
A.~Balatsoukas-Stimming and C.~Studer, ``{Deep Unfolding for Communications
  Systems: A Survey and Some New Directions},'' in \emph{\textit{Proc.} 2019
  IEEE International Workshop on Signal Processing Systems (SiPS)}, 2019, pp.
  266--271.
  
  \bibitem{8759948}
S.~Takabe, M.~Imanishi, T.~Wadayama, R.~Hayakawa, and K.~Hayashi,
``{Trainable Projected Gradient Detector for Massive Overloaded MIMO Channels: Data-Driven Tuning Approach},''
\emph{IEEE Access},  2019,vol.~7, pp. 93326--93338.

\bibitem{TISTA}
D.~Ito, S.~Takabe, and T.~Wadayama, ``{Trainable ISTA for Sparse Signal Recovery},''
\emph{in IEEE Transactions on Signal Processing}, vol. ~67, no. ~12, pp. 3113--3125, 2019.

\bibitem{He2020}
H. He, C. -K. Wen, S. Jin and G. Y. Li, "Model-Driven Deep Learning for MIMO Detection," in IEEE Transactions on Signal Processing, vol. 68, pp. 1702-1715, 2020.

\bibitem{Song}
Y.~Song and S.~Ermon,
``Generative modeling by estimating gradients of the data distribution,''
NeurIPS 2019.

\bibitem{Song2021}
Y.~Song, J.~Sohl-Dickstein, D.~P. Kingma, A.~Kumar, S.~Ermon,  and B.~Poole,
  ``Score-based generative modeling through stochastic differential
  equations,'' in Proceedings of the 10th International Conference on
  Learning Representations (ICLR), pp. 399--406, 2021.
    
\end{thebibliography}
\end{document}